\documentclass[aps,pra,10pt,twocolumn,superscriptaddress,floatfix,showpacs,numerical,footinbib,longbibliography]{revtex4-1}
\usepackage{float}       
\usepackage{amsmath}
\usepackage{amssymb}
\usepackage{dsfont}
\usepackage{bbold}
\usepackage[%
colorlinks=true,
urlcolor=blue,
linkcolor=blue,
citecolor=blue]{hyperref}
\usepackage{tikz}
\usepackage{graphicx}
\usepackage{subfigure}
\usepackage{svg}
\usepackage{mathtools}
\usepackage{relsize}

\begin{document}
	\renewcommand{\thefootnote}{\fnsymbol {footnote}}
	
	\title{\textbf{Genuine tripartite entanglement in Bhabha\\ scattering with an entangled spectator particle}}
	\author{Zan Cao}
	
	\author{Meng-Long Song}
	
	\author{Xue-Ke Song}
	
	\author{Liu Ye}
	
	\author{Dong Wang}
	\email{dwang@ahu.edu.cn}
	
	\affiliation{School of Physics, \textcolor{blue}{Anhui University}, Hefei 230601,  China}

	\begin{abstract}	
		{From the perspective of quantum information science, we investigate tree-level Bhabha scattering between an incident electron $A$ and a positron $B$, where $B$ is initially entangled with a spectator electron $C$, which does not participate in the scattering interaction.
		We find that the quantum electrodynamics (QED) scattering between $A$ and $B$ can drive the global $ABC$ system into a genuine tripartite entangled (GTE) state. Using four canonical tripartite entanglement metrics, we systematically characterize and quantify the GTE of the composite system, and demonstrate that the scattering momentum of the $A$-$B$ pair and the initial $B$-$C$ entanglement are the key resources governing GTE generation.
		We further analyze the monogamy of quantum correlations, which imposes fundamental constraints on the shareability of quantum resources in multipartite systems. Specifically, we systematically study the monogamy relations for the squared entanglement of formation and squared quantum discord in our scattering model, and find that monogamy constraints are markedly relaxed in the non-relativistic regime, enabling enhanced shareability of quantum correlations across the three particles.
		This work uncovers novel quantum correlation properties of fundamental QED scattering processes, and provides direct theoretical guidance for the development of QED-based quantum information processing protocols.}
		
	\end{abstract}
	\maketitle

	\section{Introduction}

	Quantum entanglement, the foundational cornerstone of quantum mechanics, it is the core resource enabling quantum information processing \cite{1.1QM1935,1.2QM1935,1.7RevModPhys.81.865}, and a powerful novel probe for uncovering the quantum properties of fundamental high-energy interactions \cite{1.5Pasquale2004,1.6Casini2009,1.8Book,1.9PhysRevLett.122.102001,1.10PhysRevD.97.016011}. Scattering processes – the core theoretical and experimental framework of high-energy physics, including the Bhabha scattering process studied in this work – offer a well-controlled, experimentally testable platform to study the dynamical generation, distribution, and evolution of quantum correlations in relativistic systems. Accordingly, the study of quantum entanglement in fundamental scattering processes has emerged as a vibrant interdisciplinary frontier connecting quantum information science and high-energy particle physics
	\cite {1.11JHEP,1.12PhysRevD.108.025015,1.13PhysRevC.108.L041601,1.14PhysRevD.104.116021,1.15PESCHANSKI201689,1.16PhysRevD.107.116007,1.17PhysRevLett.125.181602,1.18PhysRevD.95.114008,3.7TreelevelQRDPhysRevD2025,3.8QuarkFisher2024,3.9QuarkBell2021,3.10Bernabeu2015,3.11Lello}.

	In recent years, this interdisciplinary frontier has witnessed a series of landmark advances. On the experimental side, analyses of proton-proton collision data from the Large Hadron Collider (LHC) have reported definitive measurements of quantum entanglement and Bell nonlocality in top-antitop quark pairs \cite{2.7Afik2022quantuminformation,2.8PhysRevLett.130.221801,2.9fhkc-kfhr}; neutrino oscillation experiments have systematically verified the dynamical evolution of quantum correlations in neutrino beams  \cite{2.1Blasone20141,2.2Blasone20142,2.3Bittencourt2023,2.4PhysRevLett.117.050402,2.5PhysRevD.99.095001,2.6PhysRevA.108.032210,2.6.1PhysRevA.88.022115,2.6.2PhysRevA.98.050302,2.6.3MingFei,2.6.4WangDong,2.6.5Liyuwen,2.6.6Wangguangjie}. These results firmly establish the experimental observability of quantum correlation signals in high-energy scattering processes. On the theoretical side, extensive works have systematically characterized tree-level spin correlations in fundamental QED scattering processes, uncovering the underlying mechanism of maximal entanglement generation via elementary particle interactions \cite{2.9SciPostPhys.3.5.036,2.10Cervera2019,2.11Entanglementsaturation2025}. In particular, Bhabha scattering – the most fundamental, well-calibrated, and precisely measured process in QED – has emerged as the canonical model for studying the dynamical evolution of quantum correlations in the relativistic regime \cite{3.7TreelevelQRDPhysRevD2025}.
	A compelling extension of this research direction is the introduction of a $spectator$ $particle$, which is initially entangled with one scattering particle but does not directly participate in the scattering interaction. This configuration, first proposed in Ref. \cite{2.12spectatorPhysRevD.2019}, enables the investigation of how a single scattering event modulates entanglement between the directly interacting pair and their remote, non-interacting third particle. 
		Focusing on Bhabha scattering with an entangled spectator particle, previous studies demonstrated that such processes can mediate entanglement transfer between distinct bipartite systems, effectively acting as a quantum logic gate Refs. \cite{2.13spectatorPhysRevD.2022,2.14spectatorPhysRevD2024}. 
		However, while these pioneering works have primarily focused on bipartite correlation transfer, the dynamic generation of GTE within the global $ABC$ system during the scattering process remains largely unexplored. 
		To fill this gap, we employ the rigorous framework of quantum resource theory  \cite{3.1RevModPhys.91.025001} to investigate the tripartite correlation characteristics of the global system.
		We utilize four canonical measures of GTE: generalized geometric measure (GGM)  \cite{4.1GGMPhysRevA.81.012308, 4.2GGMPhysRevA.94.022336, 4.3GGMPhysRevA.95.022301},  three-$\pi$ entanglement \cite{4.4Threepi2007}, genuine multipartite concurrence (GMC) \cite{4.5GMC2011}, and  concurrence fill  \cite{4.6Concurrencefill}. Furthermore, monogamy is a core property of multipartite quantum correlations, governing how resources are shared among subsystems. By analyzing the monogamy relations of the squared entanglement of formation (SEF)  \cite{4.8SEFPhysRevLett.113.100503} and squared quantum discord (SQD) \cite{4.9SQDPhysRevA.88.012123}, we can achieve a more fundamental understanding of the resource distribution logic inherent in the scattering process.
	
	The core objectives of this study are threefold: (i) to analyze the impacts of the initial $B$-$C$ entanglement weight $\eta$, the scattering angle $\theta$, and the scattering momentum $\mu$ (the ratio of the incoming momentum $|\vec{p}|$ to the electron mass $m_{e}$) on the generation of GTE; (ii) to explore how relativistic effects induce characteristic evolutionary features in entanglement curves, such as entanglement saturation; and (iii) to utilize monogamy relations to evaluate the shareability of quantum resources across different momentum regimes.
	The key results of this work are summarized as follows: First, as long as the $B$-$C$ subsystem has non-zero initial entanglement, the QED scattering between $A$ and $B$ can induce the global system into a genuine tripartite entangled state. GTE exhibits non-monotonic variation with both the initial  entanglement weight $\eta$  and scattering momentum $\mu$, reaching its maximum in the intermediate parameter regime. The four tripartite entanglement measures yield fully consistent evolution laws, and concurrence fill is found to have a prominent advantage over other metrics. Second, the monogamy inequalities for the SEF and SQD hold throughout the entire scattering process, and the strength of the monogamy constraint is strongly correlated with the scattering parameters. Specifically, the monogamy constraint is markedly relaxed in the non-relativistic regime, leading to a significant enhancement in the shareability of quantum correlations among the three particles; by contrast, the monogamy constraint is extremely tightened in the relativistic limit, where both residual correlations and GTE are strongly suppressed. Third, the magnitude of residual entanglement exhibits a consistent positive correlation with the strength of GTE in the scattering-generated tripartite states, and the relaxation of monogamy constraints is a necessary prerequisite for the generation of GTE during the scattering process.
	These findings demonstrate controllable entanglement transfer and GTE generation in Bhabha scattering with an entangled spectator particle. They further establish the potential applicability of fundamental QED scattering processes for core quantum information tasks, including entanglement swapping and remote entanglement distribution \cite{EntanglementSwapping,Remote4x8dcmyx}.
	
	This paper is organized as follows. 
	In Sec. \ref{II}, we review the theoretical framework of Bhabha scattering with an entangled spectator, and then introduce several tripartite entanglement measures, together with the monogamy relations of the SEF and SQD. 
	We present the main results in Sec. \ref{III}. 
	A comprehensive discussion and concluding remarks are provided in Sec. \ref{IV}.
	\section{SETTING AND PRELIMINARIES}\label{II}
	\subsection{Model}\label{IIA}
\begin{figure}[t]
		\centering
		\includegraphics[width=0.6\linewidth]{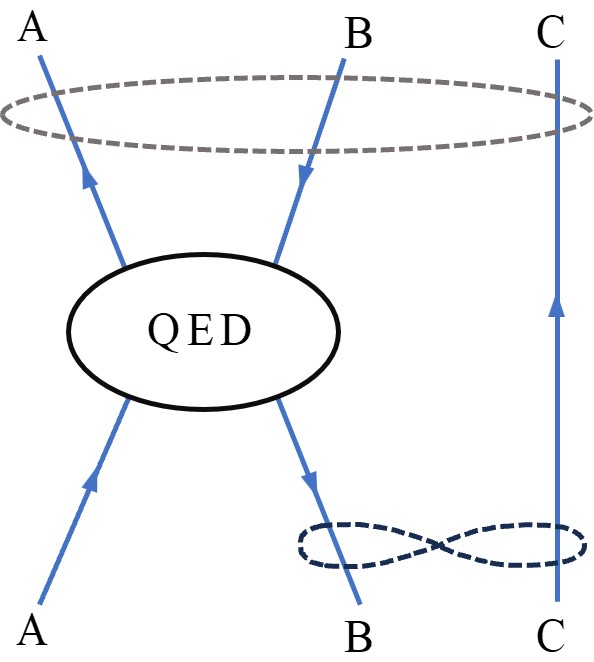}
		\caption{Consider the QED scattering process between 	electron $A$ and positron $B$. In this process, the $C$ particle is initially in an entangled state with the $B$ particle and does not directly participate in the actual scattering process.}
		\label{Fig1}
	\end{figure}

	First, we briefly review the setup used in Refs. \cite{2.12spectatorPhysRevD.2019,2.13spectatorPhysRevD.2022,2.14spectatorPhysRevD2024}, considering Bhabha scattering processes ($e^-e^+ \rightarrow e^-e^+$) in QED within the tree-level approximation, where positron $B$ is initially entangled with electron $C$, i.e., $C$ serves as the spectator particle, as schematically shown in Fig. \ref{Fig1}. For simplicity, all calculations are performed in the center-of-mass (CM) reference frame of particles $A$ and $B$, and we investigate the GTE of the global system through the $A$-$B$ scattering process. To carry out the relevant calculations, several definitions need to be introduced: in the first place, the inner product of fermionic states is defined as	
	\begin{equation}
		\langle k, a \mid p, b\rangle=2 E_{\mathbf{k}}(2 \pi)^{3} \delta^{(3)}(\mathbf{k}-\mathbf{p}) \delta_{a, b},
		\label{Eq.2.1.1}
	\end{equation}
	where $k$ and $p$ are the 4-momenta, and $a$ and $b$ are the spin indices.
	Suppose the initial state is	
	\begin{equation}
		\begin{split} 
			|i\rangle&=|p_1,a\rangle_A\otimes(\cos\eta|p_2,\uparrow\rangle_B\otimes|q,\uparrow\rangle_C\\
			&+e^{i\beta}\sin\eta|p_2,\downarrow\rangle_B\otimes|q,\downarrow\rangle_C),
			\label{Eq.2.1.2}
		\end{split}
	\end{equation}
	where  parameter $\eta$ determines the initial entanglement between $B$-$C$.  The scattering matrix (S-matrix) is defined as $S=\mathbb{I}+iT$. Here, $\mathbb{I}$ is the identity operator, which characterizes the non-scattering process.
		The operator $T$, by contrast, is the transition operator that describes the genuine Bhabha scattering process mediated by QED interactions, which constitutes the central physical process of our study.
	After scattering  the final state is
	\begin{equation}
		\begin{split}  	  
			\left|f\right\rangle
			&=|i\rangle+i\sum_{r,s}\int_{\boldsymbol{p}_3,\boldsymbol{p}_4\neq\boldsymbol{p}_1,\boldsymbol{p}_2}\delta^{(4)}(p_1+p_2-p_3-p_4) \\
			&\times[\cos\eta\mathcal{M}(a,\uparrow;r,s)|p_3,r\rangle\otimes|p_4,s\rangle\otimes|q,\uparrow\rangle \\
			&+\mathrm{e}^{i\beta}\sin\eta\mathcal{M}(a,\downarrow;r,s)|p_3,r\rangle\otimes|p_4,s\rangle\otimes|q,\downarrow\rangle],
			\label{Eq.2.1.3}
		\end{split}
	\end{equation}
	where the integral $\int_{\boldsymbol{p}} $ denotes $\int(d^3\boldsymbol{p})/(2E_p(2\pi)^3)$, $\mathcal{M}(a,\uparrow;r,s)$ represents  the scattering amplitude. Since the dependence of all $\mathcal{M}$ on the initial and final momenta is the same, for simplicity, the momentum dependence is omitted here and only the spin dependence is retained. According to Eq. (\ref{Eq.2.1.3}), the density matrix of the final state is
	\begin{equation}
		\rho_{A B C}^{f}=\frac{1}{\mathcal{N}}|f\rangle\langle f|,
		\label{Eq.2.1.4}
	\end{equation}
	where $\mathcal{N}$ is the normalization constant. Taking the partial trace over the subsystems yields
	\begin{equation}
		\text{Tr}_X[\rho_{}^f]=\sum_\sigma\int\frac{d^3\mathbf{k}}{(2\pi)^32E_\mathbf{k}}(\mathbb{I}  _r\otimes_X\langle k,\sigma|)\rho_{}^f(\mathbb{I}  _r\otimes|k,\sigma\rangle_X),
		\label{Eq.2.1.5}
	\end{equation}
	
	where, $ \mathbb{I}_r$ is the identity operator on the  remaining subspaces, $k$ and $ \sigma $ denote the 4-momentum and spin indices respectively, and $X$ defines the generic space for trace computation.
	Employing Eq. (\ref{Eq.2.1.4}) and the following relations
	\begin{equation}
		2\pi \delta^{(0)}(E_i - E_f) = \int_{-T/2}^{T/2} e^{i(E_i - E_f)t} dt,
		\label{Eq.2.1.6}
	\end{equation}
	\begin{equation}
		(2\pi)^3 \delta^{(3)}(\mathbf{k} - \mathbf{p}) = V \delta_{\mathbf{k},\mathbf{p}},
		\label{Eq.2.1.7}
	\end{equation}
	which imply that  $(2\pi) \delta^{(0)}(0) = T $ and $ (2\pi)^3 \delta^{(3)}(0) = V$, and the normalization constant $\mathcal{N}$ can be readily computed 
	\begin{equation}
		\begin{split}
			\mathcal{N} &= \text{Tr}_A [\text{Tr}_B [\text{Tr}_C [[f \rangle \langle f|]]]] \\
			&= 2E_{p_1} 2E_{p_2} 2E_q V^3 + 2E_q T^2 V^2 \Lambda,
		\end{split}
		\label{Eq.2.1.8}
	\end{equation}
	where
	\begin{equation}
		\begin{split}
			\Lambda&= \int \frac{d^3 \mathbf{p_3}}{(2\pi)^3 2 E_{p_3} 2 E_{p_3 - p_1 - p_2}} \sum_{r,s} (\cos^2 \eta |\mathcal{M}(a, \uparrow; r, s)|^2 \\
			&+ \sin^2 \eta |\mathcal{M}(a, \downarrow; r, s)|^2) |_{\mathbf{p_4} = \mathbf{p_3} - \mathbf{p_1} - \mathbf{p_2}}.
		\end{split}
		\label{Eq.2.1.9}
	\end{equation}

	\subsection{Measure of tripartite entanglement}\label{IIA}	
	
	\subsubsection{Generalized geometric measure}\label{IIA1}	
	The GGM \cite{4.1GGMPhysRevA.81.012308, 4.2GGMPhysRevA.94.022336, 4.3GGMPhysRevA.95.022301} quantifies the degree of genuine multipartite entanglement for an arbitrary $n$-partite pure state $|\psi_{N}\rangle$. It is defined as the optimal distance between the given state and the set of non-genuine multipartite entangled states. Its explicit form is given below
	\begin{equation}
		\mathcal{G}(|\psi_{N}\rangle)=1-\Lambda_{max}^{2}(|\psi_{N}\rangle),
		\label{Eq.2.2.1}
	\end{equation}
	where
	\begin{equation}
		\Lambda_{\max }\left(\left|\psi_{N}\right\rangle\right)=\max \left|\left\langle\chi \mid \psi_{N}\right\rangle\right|=\max _{|\chi\rangle} F\left(\left|\psi_{N}\right\rangle,|\chi\rangle\right).
		\label{Eq.2.2.2}
	\end{equation}
	This maximization is carried out over all non-genuine multipartite entangled states $|\chi\rangle$, where $F\left(\left|\psi_{N}\right\rangle,|\chi\rangle\right)$ represents the fidelity between $\left|\psi_{N}\right\rangle$  and $|\chi\rangle$. By employing the Hilbert-Schmidt distance, we can recast Eq. (\ref{Eq.2.2.2}) into the equivalent form below
	\begin{equation}
		\mathcal{G}(|\psi\rangle)=1-\max\left\{\varepsilon_{I:L}^2|I\cup L=\{1,2,\ldots,n\},I\cap L=\emptyset\right\},
		\label{Eq.2.2.3}
	\end{equation}
	where $\mathcal{E}_{I:L}$ denotes the maximum Schmidt coefficient of $\left|\psi_{N}\right\rangle$ for the bipartition $ I : L$. For arbitrary pure states,  $\mathcal{E}_{I:L}^2$ equals the corresponding eigenvalues of the reduced density matrices $\rho_{I}$ and $\rho_{L}$.

	\subsubsection{Three-$\pi$ entanglement}\label{IIA2}	
	For a tripartite pure state $\left|\psi\right\rangle_{ABC}$, the Coffman–Kundu–Wootters (CKW)-like monogamy inequality quantified via negativity \cite{5.1GMIPhysRevLett.96.220503} is given by
	\begin{equation}
		N_{AB}^{2}+N_{AC}^{2}\leq N_{A(BC)}^{2},
		\label{Eq.2.2.4}
	\end{equation}
	where $N_{AB}$ and $N_{AC}$ denote the sum of the negative eigenvalues of the partial transpose matrices of the reduced states $\rho_{AB}=\mathrm{Tr}_C(\rho_{ABC})$ and $\rho_{AC}=\mathrm{Tr}_B(\rho_{ABC})$, respectively. Following the approach outlined in Ref. \cite{4.4Threepi2007}, we have
	\begin{equation}
		N_{A(BC)}=C_{A(BC)}=\sqrt{2[1-\mathrm{Tr}(\rho_{A}^{2})]},
		\label{Eq.2.2.5}
	\end{equation}
	where $\rho_{A} =\mathrm{Tr}_{BC}(\rho_{ABC})$. $C_{A(BC)}$ is defined as the one-versus-rest bipartite entanglement, i.e., the entanglement between qubit $A$ and the composite subsystem formed by the other two qubits ($B$ and $C$). 
	Residual entanglement is further defined as the difference between the two sides of Eq. (\ref{Eq.2.2.4})
	\begin{equation}
		\pi_{A}=N_{A(BC)}^{2}-N_{AB}^{2}-N_{AC}^{2}.
		\label{Eq.2.2.6}
	\end{equation}
	Similarly, focusing on subsystems $B$ and $C$, two additional types of residual entanglement can be generated
	\begin{equation}
		\begin{split}
			&\pi_{B}=N_{B(AC)}^{2}-N_{BA}^{2}-N_{BC}^{2},\\
			&\pi_{C}=N_{C(AB)}^{2}-N_{CA}^{2}-N_{CB}^{2}.
			\label{Eq.2.2.7}
		\end{split}
	\end{equation}
	Note that the residual entanglement for different focal subsystems varies under qubit transformations. Lastly, the three-$\pi$ entanglement of the tripartite system is defined as the average of $\pi_{A}$, $\pi_{B}$, and $\pi_{C}$
	\begin{equation}
		\pi_{ABC}=\frac{1}{3}(\pi_{A}+\pi_{B}+\pi_{C}).
		\label{Eq.2.2.8}
	\end{equation}

	\subsubsection{Genuinely multipartite concurrence}\label{IIA3}
	The GMC is a computable measure for quantifying multipartite entanglement, based on the well-known concurrence \cite{4.5GMC2011}. For an $n$-partite pure state $|\Psi\rangle\in\mathcal{H}_{1}\otimes\mathcal{H}_{2}\otimes\cdots\otimes\mathcal{H}_{\mathrm{n}}$, the GMC is defined as
	\begin{equation}
		C_{\text{GMC}}(|\Psi\rangle)=\min_{\gamma_{i}\in\gamma}\sqrt{2[1-\mathrm{Tr}(\rho_{A_{\gamma_{i}}}^{2})]}
		\label{Eq.2.2.9}
	\end{equation}
	where $\gamma=\{\gamma_{\mathrm{i}}\}$ represents the entire set of all possible bipartitions $\{A_i | B_i\}$ of the set $\{1, 2,..., n\}$. The GMC can be further extended to the mixed-state scenario via the convex roof construction
	\begin{equation}
		C_{\text{\text{GMC}}}(\rho)=\operatorname*{inf}_{\{p_{i},|\psi_{i}\rangle\}}\sum_{i}p_{i}C_{\text{GMC}}(|\psi_{i}\rangle),
		\label{Eq.2.2.10}
	\end{equation}
	where the infimum is taken over all possible convex decompositions of $\rho$ as $\rho=\sum_{i}\left|\psi_{i}\right\rangle\left\langle\psi_{i}\right|$.
	
	\subsubsection{Concurrence fill}\label{IIA4}
	For tripartite entangled states, concurrence fill was introduced as a robust entanglement measure based on the entanglement triangle area method \cite{4.6Concurrencefill}. In this formulation, the side lengths of the triangle are equal to the squares of the three bipartite concurrences. With Heron’s triangle area formula, the concurrence fill is  defined as
	
	\begin{equation}
		\begin{split}
			&\mathcal{F}(|\psi\rangle)=\\
			&\left[\frac{16}{3}Q\left(Q-C_{A(BC)}^{2}\right)\left(Q-C_{B(AC)}^{2}\right)\left(Q-C_{C(AB)}^{2}\right)\right]^{1/4},
		\end{split}\label{Eq.2.2.11}
	\end{equation}
	where, $Q=\frac{1}{2}\left(C_{A(BC)}^{2}+C_{B(AC)}^{2}+C_{C(AB)}^{2}\right)$ denotes the semiperimeter of the entanglement triangle, which corresponds to the global entanglement degree \cite{5.2Globalentanglement,5.3Anobservablemeasure}. 
	The factor 16/3 acts as the normalization factor to ensure $0\leqslant \mathcal{F}\leqslant1$, and the additional square-root term in addition to Heron’s formula ensures local monotonicity under LOCC (Local Operations and Classical Communication). The bipartite concurrence $C_{i(jk)}$ is calculated as follows  \cite{5.4PhysRevA.61.052306}
	\begin{equation}
		C_{i(jk)}=2\sqrt{\det\rho_{i}},
		\label{Eq.2.2.12}
	\end{equation}
	where $i,j,k\in\{A,B,C\}$ are mutually distinct, and $\rho_{i}$  represents the reduced density matrix associated with the global state $\rho_{A B C}$. We observe that $C_{i(jk)}$ obeys the inequality $0\leq C_{i(jk)}\leq1$.

	\subsection{Monogamy relation of quantum correlations}\label{IID}
	\subsubsection{Monogamy relation for SEF}\label{IID1}	
	
	For an arbitrary two-qubit state $\rho_{AB}$, Wootters \cite{6.1PhysRevLett.80.2245} derived an explicit analytical expression for the entanglement of formation.
	\begin{equation}
		E_{f}\left(\rho_{AB}\right)=H\left(\frac{1+\sqrt{1-\left|C(\rho_{AB})\right|^{2}}}{2}\right),
		\label{Eq.2.2.15}
	\end{equation}
	where $H(x)=-x\log_2x-(1-x)\log_2(1-x)$ is the binary  entropy, $C(\rho_{AB})=\max \left\{\sqrt{\lambda_1}-\sqrt{\lambda_2}-\sqrt{\lambda_3}-\sqrt{\lambda_4},0\right\}$ denotes the concurrence of the density matrix $\rho_{AB}$. Here, $\lambda_i$ are the eigenvalues of   $\rho_{AB}\widetilde{\rho}_{AB}$ with decreasing order, with $\widetilde{\rho}_{AB}= \begin{pmatrix} \sigma_y\otimes\sigma_y \end{pmatrix}\rho_{AB}^* \begin{pmatrix} \sigma_y\otimes\sigma_y \end{pmatrix}$.
	For any three-qubit state $\rho_{A B C}$, the bipartite entanglement measured by SEF satisfies the following pairwise inequality
	\cite{4.8SEFPhysRevLett.113.100503}
	\begin{equation}
		E_{f}^{2}(\rho_{A|BC})\geq E_{f}^{2}(\rho_{AB})+E_{f}^{2}(\rho_{AC}),
		\label{Eq.2.2.16}
	\end{equation}
	where  $E_{f}^{2}(\rho_{A|BC})$ quantifies the entanglement between subsystem $A$ and the composite $BC$, whereas $E_{f}^{2}(\rho_{AB})$ and $E_{f}^{2}(\rho_{AC})$ represent the pairwise entanglement between $A$ and $B$ and between $A$ and $C$, respectively.

	\subsubsection{Monogamy relation for SQD }\label{IID2}	
	Apart from entanglement, quantum discord (QD) is another pivotal measure of bipartite quantum correlation, with its formal definition provided in Refs. \cite{6.2PhysRevLett.88.017901,6.3LHenderson2001}
	\begin{equation}
		D(\rho_{AB})=\widetilde{S}(\rho_A|\rho_B)-S(\rho_A|\rho_B),
		\label{Eq.2.2.17}
	\end{equation}
	where $\widetilde{S}(\rho_{A}|\rho_{B})=\min_{\left\{M_{j}^{B}\right\}}\sum_{j}p_{j}S\left(\rho_{A|j}\right)$ denotes the measurement-induced quantum conditional entropy. Here, $\{M_{j}^{B}\}$ is a positive operator-valued measure applied to subsystem $B$, and $S(\rho_A|\rho_B)=S(\rho_{AB} )-S(\rho_B)$ is the conditional entropy of $A$ given $B$. Particularly, for a tripartite pure state $|\psi_{ABC}\rangle$, the pairwise quantum discord can be derived by combining Eq. (\ref{Eq.2.2.17}) with the Koashi–Winter formula \cite{6.4PhysRevA.69.022309}
	\begin{equation}
		D(\rho_{ik})=E_{f}(\rho_{ij})-S(\rho_{i}|\rho_{k}).
		\label{Eq.2.2.18}
	\end{equation}
	The measurement is performed on subsystem $k$, with $i,j,k\in{A,B,C}$ are mutually distinct. Moreover, a well-defined relationship exists between QD and the entanglement of formation \cite{6.2PhysRevLett.88.017901,6.3LHenderson2001}
	\begin{equation}
		D\left(\rho_{i|jk}\right)=E\left(\rho_{i|jk}\right)=S\left(\rho_{i}\right).
		\label{Eq.2.2.19}
	\end{equation}
	For an arbitrary three-qubit pure state $\rho_{ABC}$, the bipartite quantum correlation quantified by SQD satisfies the monogamy inequality reported in Ref. \cite{4.9SQDPhysRevA.88.012123}
	\begin{equation}
		D^{2}(\rho_{A|BC})\geq D^{2}(\rho_{AB})+D^{2}(\rho_{AC}).
		\label{Eq.2.2.20} 
	\end{equation}

	\begin{figure*}[t]
		\centering
		\includegraphics[width=17.8cm]{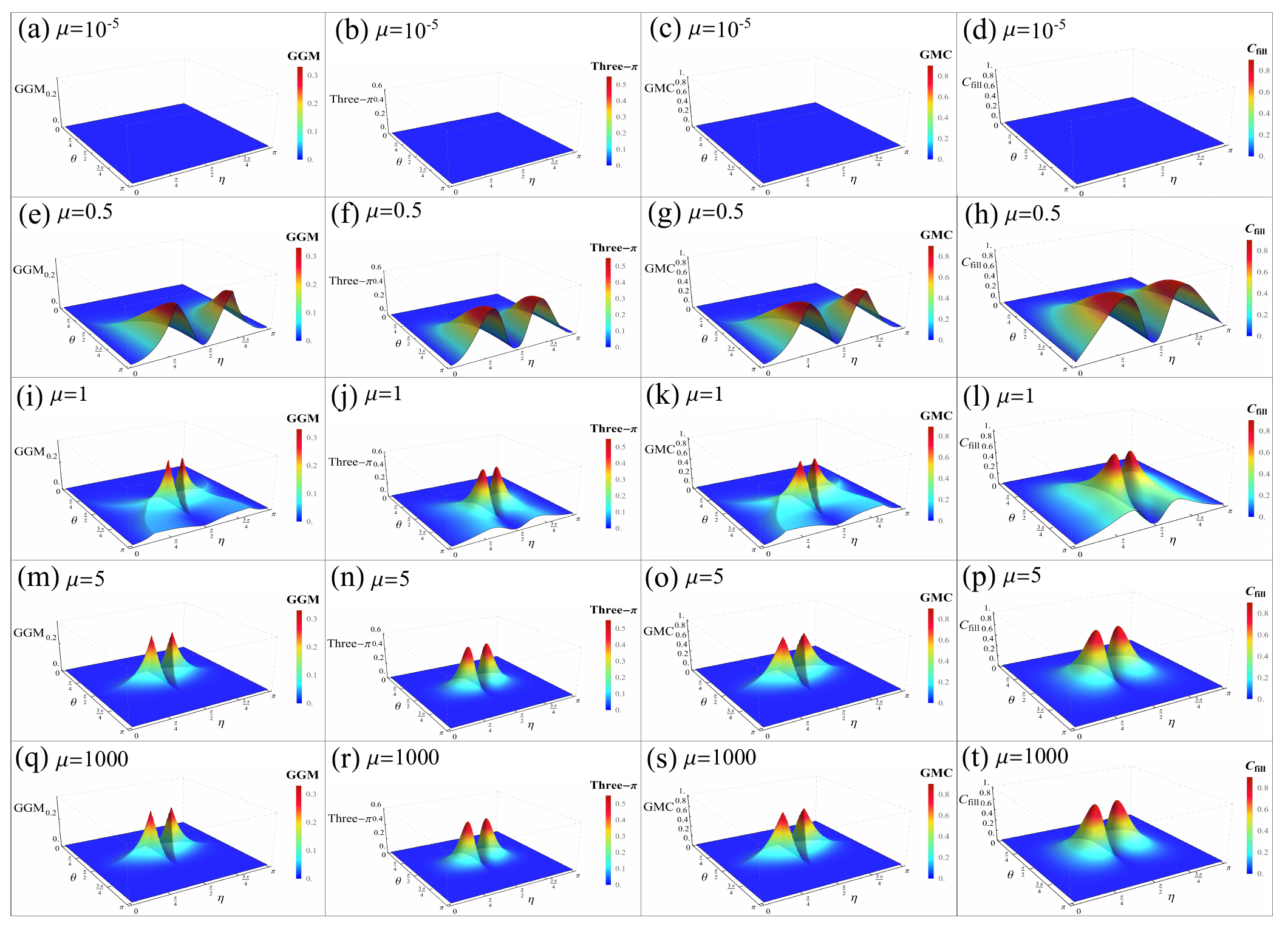}
		\caption{The GTE is plotted against the scattering angle $\theta$ and initial entanglement weight $\eta$ for distinct values of the scattering momentum $\mu$.
				Herein, panels (a)–(d) correspond to $\mu=10^{-5}$(nonrelativistic limit). Panels (e)–(h) to $\mu=0.5$, panels (i)–(l) to $\mu=1$ (non-relativistic regime). Panels (m)–(p) to $\mu=5$, and panels (q)–(t) to $\mu =1000$ (relativistic regime). Each column of panels corresponds to one of the entanglement measures in sequence: GGM, Three-$\pi$ entanglement, GMC, and concurrence fill.}
		\label{Fig2}
	\end{figure*}	
	\section{results and analysis}\label{III}	
	\subsection{Scattering generated genuine tripartite entanglement}\label{IIIA}	
	Assuming that the initial state of the system $ABC$ is
	\begin{equation}
		|i\rangle=|R\rangle_{A}\otimes(\cos\eta|R\rangle_{B}|R\rangle_{C}+e^{i\beta}\sin\eta|L\rangle_{B}|L\rangle_{C}),
		\label{Eq.3.1.1}
	\end{equation}
	accordingly, its density matrix can be expressed as
	\begin{equation}
		\begin{aligned}
			\rho_{ABC}^{i} & =\mathrm{cos}^{2}\eta|R\rangle_{A}|R\rangle_{B}|R\rangle_{CC}\langle R|_{B}\langle R|_{A}\langle R|\\
			&+e^{-i\beta}\sin\eta\cos\eta|R\rangle_{A}|R\rangle_{B}|R\rangle_{CC}\langle L|_{B}\langle L|_{A}\langle R| \\
			& +e^{i\beta}\sin\eta\cos\eta|R\rangle_{A}|L\rangle_{B}|L\rangle_{CC}\langle R|_{B}\langle R|_{A}\langle R|\\
			&+\mathrm{sin}^{2}\eta|R\rangle_{A}|L\rangle_{B}|L\rangle_{CC}\langle L|_{B}\langle L|_{A}\langle R|.
		\end{aligned}
		\label{Eq.3.1.2}
	\end{equation}
In the CM reference frame of particles $A$ and $B$, the incoming 4-momenta  lie along the $z$-axis denoted as 
		\begin{equation}
			\begin{split}
				&p_1 = (\omega, 0, 0, |\vec{p}|),\\
				&p_2 = (\omega, 0, 0, -|\vec{p}|),
				\label{Eq.55}
			\end{split}
		\end{equation}		
		after the interaction, the outgoing 4-momenta are
		\begin{equation}
			\begin{split}
				&p_3 = (\omega, |\vec{p}|\sin\theta, 0, |\vec{p}|\cos\theta),\\
				&p_4 = (\omega, -|\vec{p}|\sin\theta, 0, -|\vec{p}|\cos\theta),
				\label{Eq.66}
			\end{split}
		\end{equation}	
		where the scattering angle $\theta$ is defined with respect to the $z$-axis. For convenience in characterizing the scattering energy, define the dimensionless scattering momentum, $\mu=|\mathbf{p}|/m_e$, $|\mathbf{p}|$ is the incoming momentum in the CM reference frame and ${m_e}$ is the electron mass.
	According to Eqs. (\ref{Eq.2.1.3}) and (\ref{Eq.2.1.4}), the density matrix of the full final state can be derived. However, since we are specifically interested in the entanglement properties generated by actual scattering events, we focus on the particles scattered at non-zero angles. By projecting out the unscattered forward beam (which corresponds to the identity part of the S-matrix), the post-scattering density matrix, restricted solely to the interaction part, is given by:
	\begin{figure*}[htbp]
		\centering
		\includegraphics[width=17.5cm]{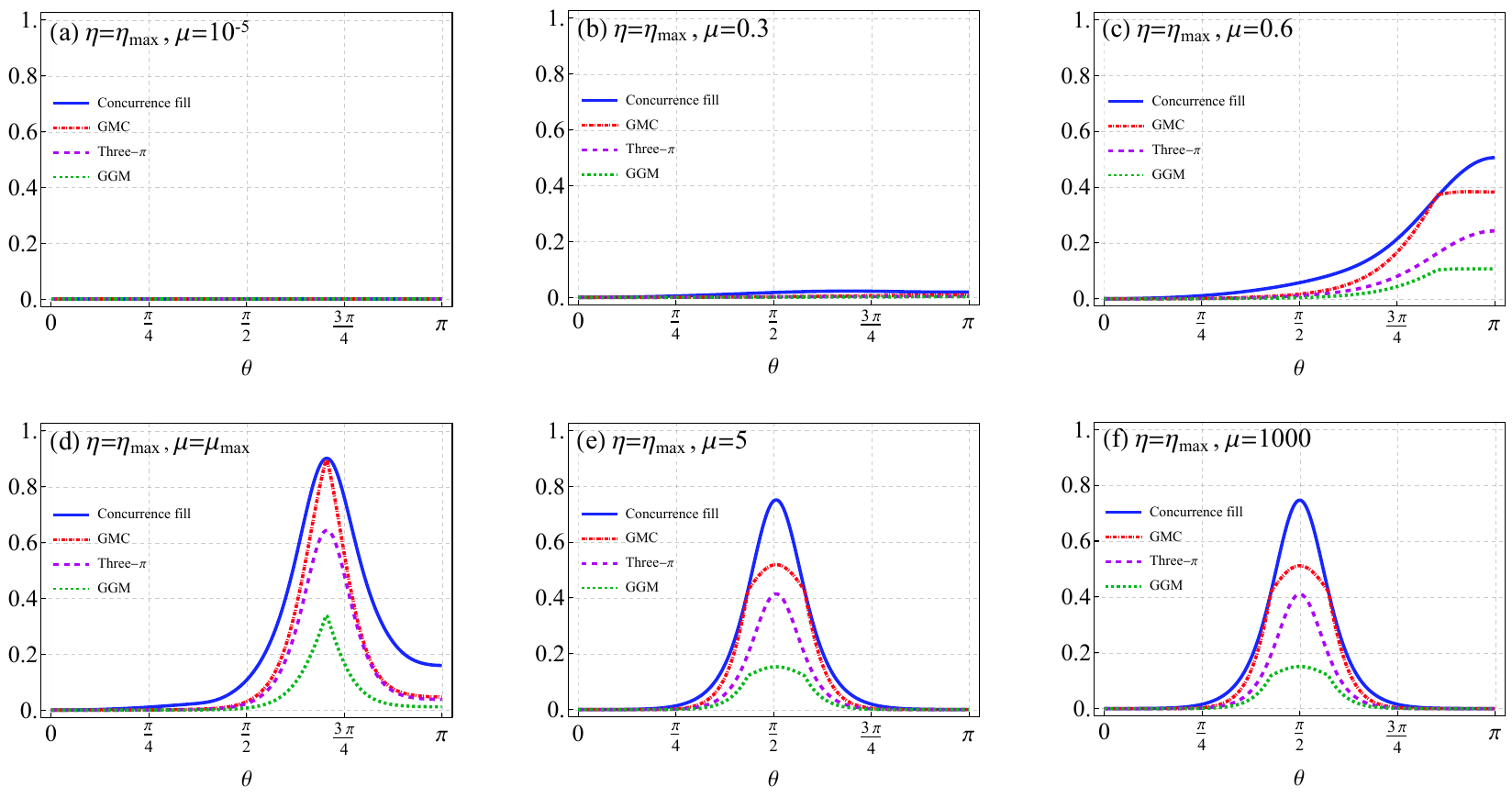}
		\caption{
			The GTE is plotted against the scattering angle $\theta$, with $\eta$ fixed at $\eta_{\rm max}$ for several distinct values of the scattering momentum $\mu$.
			Solid blue represents  Concurrence  fill, dashed red represents  GMC, dash-dotted purple denotes Three-$\pi$ entanglement, dotted green denotes  GGM.}
		\label{Fig3}
	\end{figure*}
	\begin{widetext}
		\begin{equation}\begin{aligned}
				\rho_{ABC}^{f} & =\sum_{r,s,r^{\prime},s^{\prime}}\left[\mathrm{cos}^{2}\eta\mathcal{M}(RR,rs)\mathcal{M}^{\dagger}(RR,r^{\prime}s^{\prime})|r\rangle_{A}|s\rangle_{B}|R\rangle_{CC}\langle R|_{B}\langle s^{\prime}|_{A}\langle r^{\prime}|\right. \\
				& +e^{-i\beta}\sin\eta\cos\eta\mathcal{M}(RR,rs)\mathcal{M}^\dagger(RL,r^{\prime}s^{\prime})|r\rangle_A|s\rangle_B|R\rangle_{CC}\langle L|_B\langle s^{\prime}|_A\langle r^{\prime}| \\
				& +e^{i\beta}\sin\eta\cos\eta\mathcal{M}(RL,rs)\mathcal{M}^\dagger(RR,r^\prime s^\prime)|r\rangle_A|s\rangle_B|L\rangle_{CC}\langle R|_B\langle s^\prime|_A\langle r^\prime| \\
				& +\sin^2\eta\mathcal{M}(RL,rs)\mathcal{M}^\dagger(RL,r^\prime s^\prime)|r\rangle_A|s\rangle_B|L\rangle_{CC}\langle L|_B\langle s^\prime|_A\langle r^\prime|\Big],
				\label{Eq.3.1.3}
			\end{aligned}
		\end{equation}
	\end{widetext}
	where, $\mathcal{M}$ are scattering amplitudes as functions of  $\theta$ and $\mu$, the explicit forms of which are given in Appendix \ref {A}.
	
	By combining Eqs. (\ref{Eq.3.1.2})-(\ref{Eq.3.1.3}), we can derive the reduced density matrices  $\rho_A^{i(f)}$, $\rho_B^{i(f)}$, $\rho_C^{i(f)}$ and $\rho_{AB}^{i(f)}$, $\rho_{AC}^{i(f)}$, $\rho_{BC}^{i(f)}$, whose explicit forms are reported in Appendix \ref{B}.
	Using the density matrix of the final state, the GGM can  be evaluated via Eq. (\ref{Eq.2.2.3}). 
	\begin{equation}
		\text{GGM}(\rho_{ABC}^f)=1-\max\left \{\lambda_A,\lambda_B,\lambda_C \right \}, 
	\end{equation}
	where $\lambda_A$, $\lambda_B$ and $\lambda_C$  denote the maximum eigenvalue of the reduced density matrices $\rho_A$, $\rho_B$ and $\rho_C$, respectively. To calculate three-$\pi$ entanglement, we first take subsystem $A$ as the reference to perform the partial transpose operation on $\rho_{AB}$, $\rho_{AC}$, yielding $\rho_{AB}^T$ and $\rho_{AC}^T$. By extracting the negative eigenvalues of these transposed matrices, we can compute the negativities $N_{AB}$ and $N_{AC}$. Subsequently, the value of three-$\pi$ entanglement is obtained using Eqs. (\ref{Eq.2.2.6})-(\ref{Eq.2.2.8}). According to Eq. (\ref{Eq.2.2.9}), the GMC is given by min$\left\{2(1-\mathrm{Tr}(\rho_{A}^{2})),2(1-\mathrm{Tr}(\rho_{B}^{2})),2(1-\mathrm{Tr}(\rho_{C}^{2}))\right\}$. Finally, the concurrence fill is determined by calculating the area of the concurrence triangle via Eqs. (\ref{Eq.2.2.11}) and (\ref{Eq.2.2.12}).
	
	Through analytical calculations, we derive the closed-form expression for the concurrence fill in the relativistic limit, 
	\begin{equation}
		C_\text{fill}=\lim_{\mu \to \infty} \frac{256(\frac{A \sin^{6}{\eta} \sin^{4}{2\eta}\sin^{16}{\theta}}{B^{8}} )^{1/4}}{3^{1/4}}, 
		\label{Eq.2.2.30}
	\end{equation}
	where
\begin{equation}
	\begin{aligned}
		A&=17955+14280\cos{2\theta}+540\cos{4\theta}-8\cos{6\theta}\\
		&+\cos{2\eta}(17885+14392\cos{2\theta}+484\cos{4\theta}\\
		&+8\cos{6\theta}-\cos{8\theta})+\cos{8\theta}, \\
		B &= 99 + 29\cos{2\eta} + 2(28\cos{2\theta} + \cos{4\theta})\sin^{2}{\eta}.
	\end{aligned}
	\label{Eq.3.1.266}
\end{equation}
	For other cases, only the numerical results of the GTE can be obtained. It is important to note that while the relative phase parameter $\beta$ is present in the off-diagonal elements of the density matrices, all subsequent calculations of quantum correlations (GTE, SEF, and SQD) depend exclusively on the eigenvalues of these matrices or their specific transformations. Consequently, the phase factor $e^{i\beta}$ cancels out during the eigenvalue evaluations, rendering all entanglement and discord measures evaluated in this study strictly independent of $\beta$.
	\begin{figure*}[htbp]
		\centering
		\includegraphics[width=17.5cm]{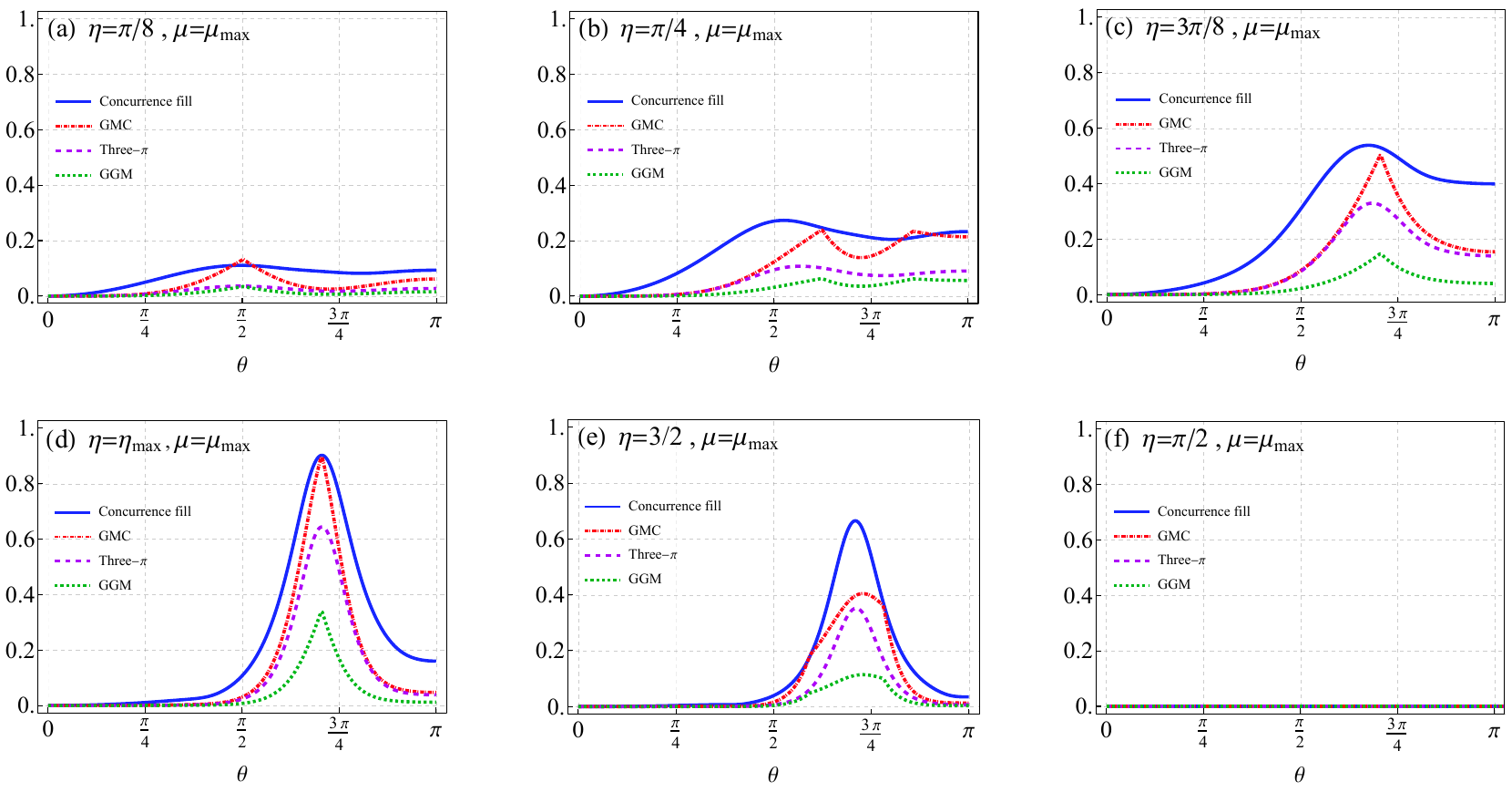}
		\caption{The GTE is plotted against the scattering angle $\theta$, with the scattering momentum $\mu$ fixed at $\mu_{\rm max}$, for several distinct values of $\eta$.
			Solid blue represents  Concurrence  fill, dashed red represents  GMC, dash-dotted purple denotes Three-$\pi$ entanglement, dotted green denotes  GGM.}
		\label{Fig4}
	\end{figure*}
	
	Fig. \ref{Fig2} displays the three-dimensional profiles of GTE for the global system, plotted against the scattering angle $\theta$ and initial entanglement weight $\eta$ with different values of the scattering momentum  $\mu$. Our findings indicate that in the non-relativistic limit ($\mu=10^{-5}$), GTE vanishes regardless of the scattering angle $\theta$ and the initial entanglement weight $\eta$. In contrast, $A$-$B$ scattering induces GTE within both the non-relativistic ($\mu=0.5,1$) and relativistic regimes ($\mu=5,1000$). All four entanglement measures provide consistent and effective quantifications, confirming that the scattering momentum $\mu$ serves as a pivotal resource for the generation of GTE.
	{A more detailed analysis shows that  when $\eta = 0$ 
		(i.e., subsystems $B$ and $C$ are in a product state), no GTE can be generated, irrespective of the scattering angle $\theta$ and scattering momentum $\mu$.}
	{For specific scattering angles, the GTE increases monotonically with increasing $\eta$ at first, subsequently, as $\eta$ reaches a certain intermediate value, the GTE gradually decreases, eventually dropping to zero  at $\eta=\pi/2$.}
	Notably, the GTE does not reach its maximum when $\eta=\pi/4$, a condition corresponding to a maximally entangled $B$-$C$ state. This indicates that the GTE of the global system does not exhibit a monotonic dependence on the initial entanglement of the $B$-$C$ subsystem. Collectively, these results demonstrate that the initial $B$-$C$ entanglement is a necessary resource for GTE generation in the global system, whereas a maximally entangled $B$-$C$ state suppresses the formation of the peak GTE. 
	This phenomenon can be explained by the entanglement monogamy relationship described in the Sec. \ref{IIIB}.
	Furthermore, the plots indicate that for the same entanglement measure, non-vanishing GTE can be detected across a broader parameter range in the non-relativistic regime, as exemplified by Figs. \ref{Fig2}\textcolor{blue}{(a)}, \textcolor{blue}{(e)}, \textcolor{blue}{(i)}, \textcolor{blue}{(m)}. Numerical calculations reveal that the four tripartite entanglement measures attain their peak values at nearly identical parameter configurations ($\theta_{\rm max}\approx2.21, \eta _{\rm max}\approx1.42, \mu _{\rm max}\approx0.913$), with the corresponding peak values being approximately 0.342 for GGM, 0.648 for Three-$\pi$, 0.900 for GMC, and 0.902 for concurrence fill.
	Although these four GTE measures are derived from distinct theoretical approaches—ranging from distance bounds to geometric areas—they exhibit strong phenomenological relations in this scattering model. Specifically, they provide consistent qualitative mappings of the entanglement dynamics, universally peaking at identical parameter configurations.
	
	Fig. \ref{Fig3} illustrates the variation of the GTE with the scattering angle $\theta$  for distinct values of the scattering momentum $\mu$, with $\eta$ fixed at $\eta_{\rm max}$.
	Initially, the GTE rises with increasing $\mu$; subsequently, once $\mu$ exceeds the critical value $\mu=\mu_{\rm max}$, the GTE decreases gradually; ultimately, as $\mu$ progresses to the relativistic limit, the GTE converges to a saturation value. These results indicate that the scattering momentum $\mu$ is a necessary condition for the generation of GTE. We further note that the formation of peak GTE is suppressed in both the non-relativistic regime ($\mu=0.3,0.6$) and the relativistic limit ($\mu=1000$) of the $A$-$B$ scattering process.
	Furthermore, in the non-relativistic regime, the GTE reaches its maximum at $\theta=\pi$; in the relativistic limit, GTE peaks emerge at  $\theta=\pi/2$. Notably, under identical parameter settings, the four GTE measures satisfy the following hierarchical relationship
	\begin{figure*}[htbp]
		\centering
		\includegraphics[width=17.5cm]{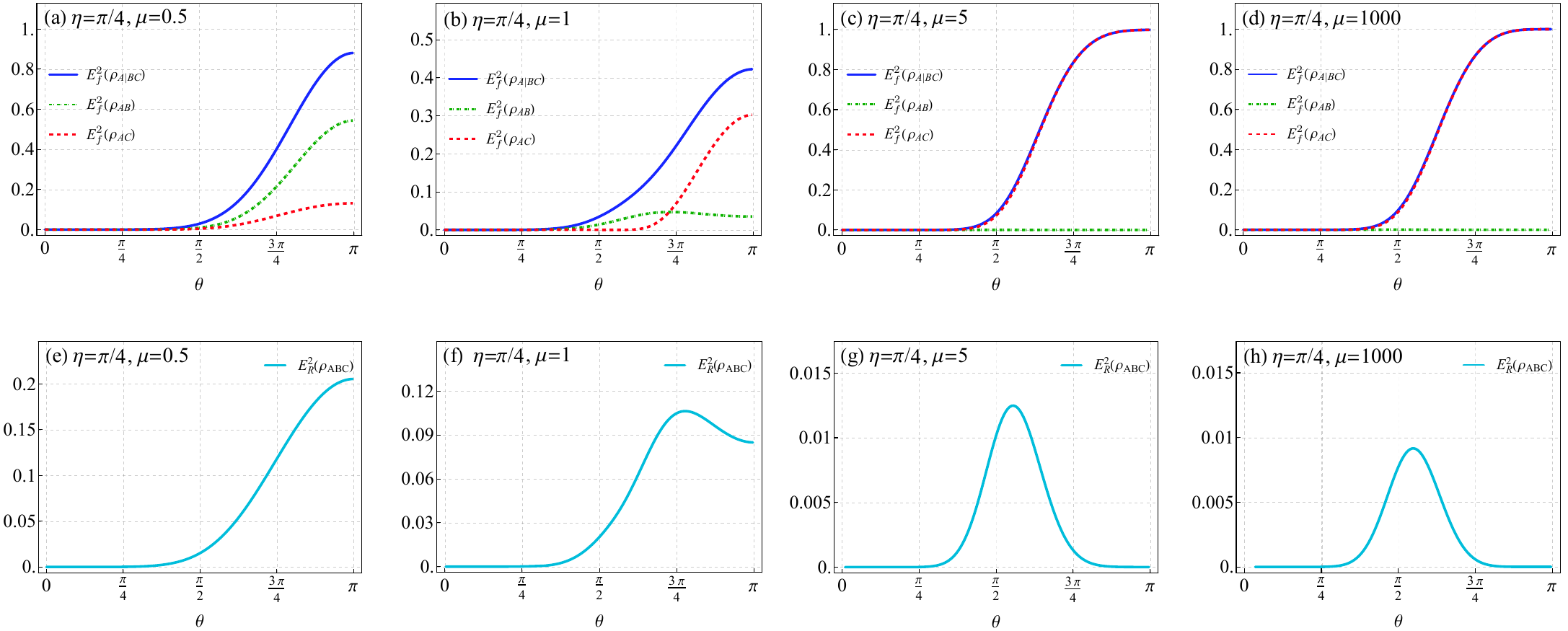}
		\caption{	
			The entanglement monogamy relation is shown against the scattering angle $\theta$ for various values of the scattering momentum $\mu$, where $\eta$ is set to $\pi/4$.		
			Herein, the solid blue curve in panels (a)–(d) represents $ E_{f}^{2}(\rho_{A|BC})$, the green dash-dotted curve represents $ E_{f}^{2}(\rho_{AB})$, the red dashed curve represents $ E_{f}^{2}(\rho_{AC})$;
			Panels (e)–(h) plot the  $E_R^{2}\left(\rho_{ABC}\right)$ as a function of the scattering angle $\theta$.}
		\label{Fig5}
	\end{figure*}
	\begin{equation}
		\text{Concurrence fill}\ge \text{GMC} \ge \text{Three-}\pi \ge \text{GGM}.
	\end{equation}
	While this specific ordering is characteristic of the states generated in this process rather than a strictly universal inequality for all tripartite systems, the consistently higher value of the concurrence fill is noteworthy. Unlike the GGM or GMC, which establish bounds based on singular extremal values (such as maximum eigenvalues or minimum bipartition concurrences) , the concurrence fill is derived from the area of the entanglement triangle. By utilizing the semiperimeter $Q$ alongside all three bipartite concurrences, it analytically integrates the global entanglement degree and all internal pairwise correlations into a single metric, providing a highly comprehensive geometric quantification of the system's genuine tripartite entanglement.
	
	In Fig. \ref{Fig4}, we fix the scattering momentum at $\mu=\mu_{\rm max}$ to further investigate the evolution of the GTE with the scattering angle $\theta$ for different values of $\eta$.
	Initially, the GTE increases gradually with rising  $\eta$; once  $\eta$ exceeds the critical value $\eta _{\rm max}$, the GTE declines rapidly and ultimately drops to zero at $\eta=\pi/2$. We find that for small $\eta$, the GGM exhibits marginally better performance than the concurrence fill metric in quantifying GTE at specific scattering angles, as illustrated in Figs.  \ref{Fig4}\textcolor{blue}{(a)-}\ref{Fig4}\textcolor{blue}{(b)}. From a global perspective, however, the concurrence fill metric demonstrates a distinct advantage in  the quantification of GTE.
	It can be seen that GMC and GGM exhibit non-analytical sharp peaks due to the nonanalytic minimum argument in their expressions, whereas the concurrence fill and three-$\pi$ are always smooth.

	\subsection{Monogamy relations for SEF}\label{IIIB}	
	To examine the monogamy of entanglement quantified by  SEF in this scattering model, the residual entanglement is defined as
	\begin{equation}
		E_{R}^{2}\left(\rho_{ABC}\right)=E_{f}^{2}\left(\rho_{A|BC}\right)-E_{f}^{2}\left(\rho_{AB}\right)-E_{f}^{2}\left(\rho_{AC}\right),
	\end{equation}
	\begin{figure}[htbp]
		\centering
		\includegraphics[width=7cm]{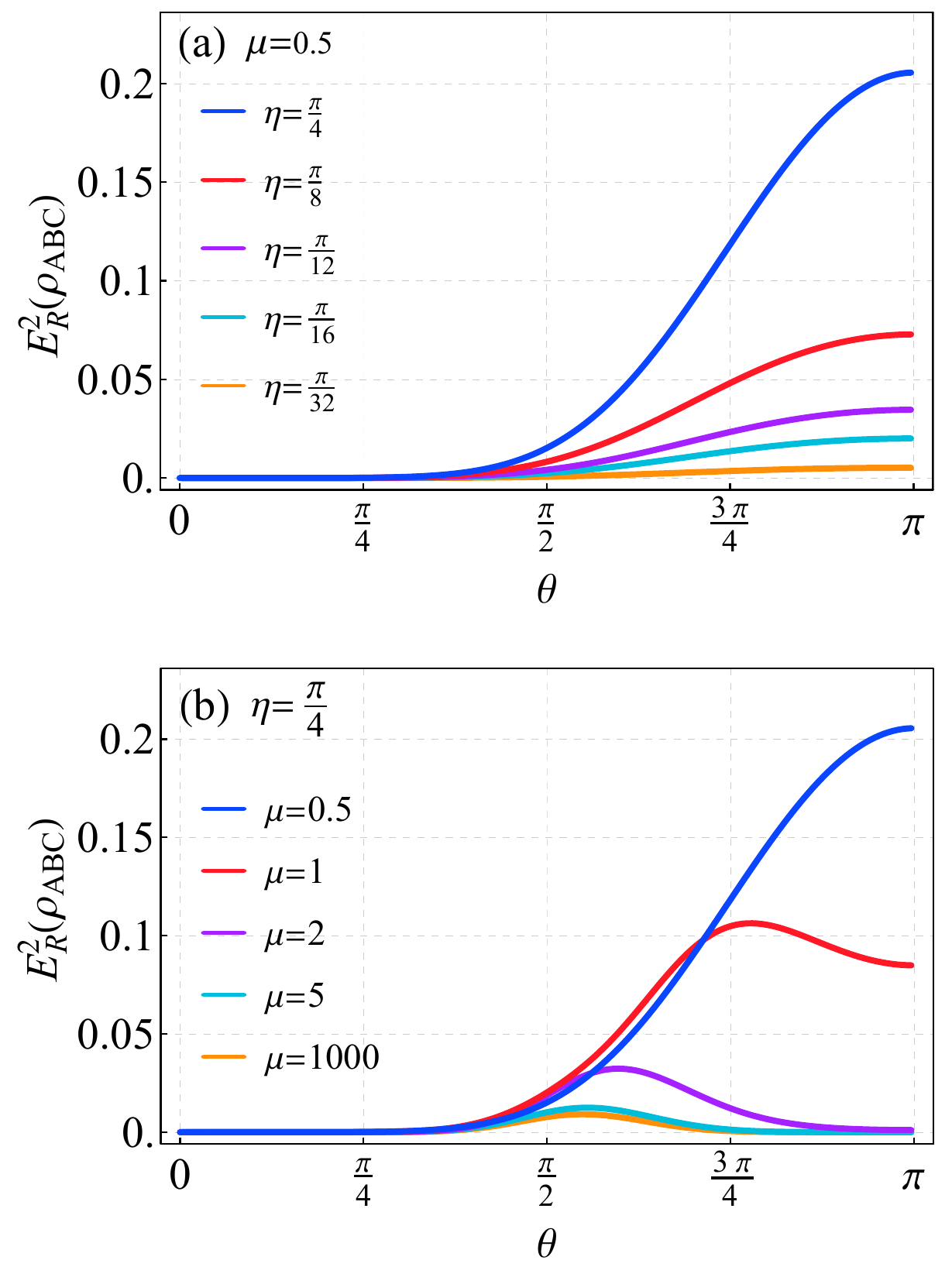}
		\caption{Residual entanglement is plotted as a function of the  scattering angle  $\theta$ for specific scattering momentum $\mu$ and initial entanglement weight $\eta$.}
		\label{Fig6}
	\end{figure}
	\begin{figure*}[htbp]
		\centering
		\includegraphics[width=17.5cm]{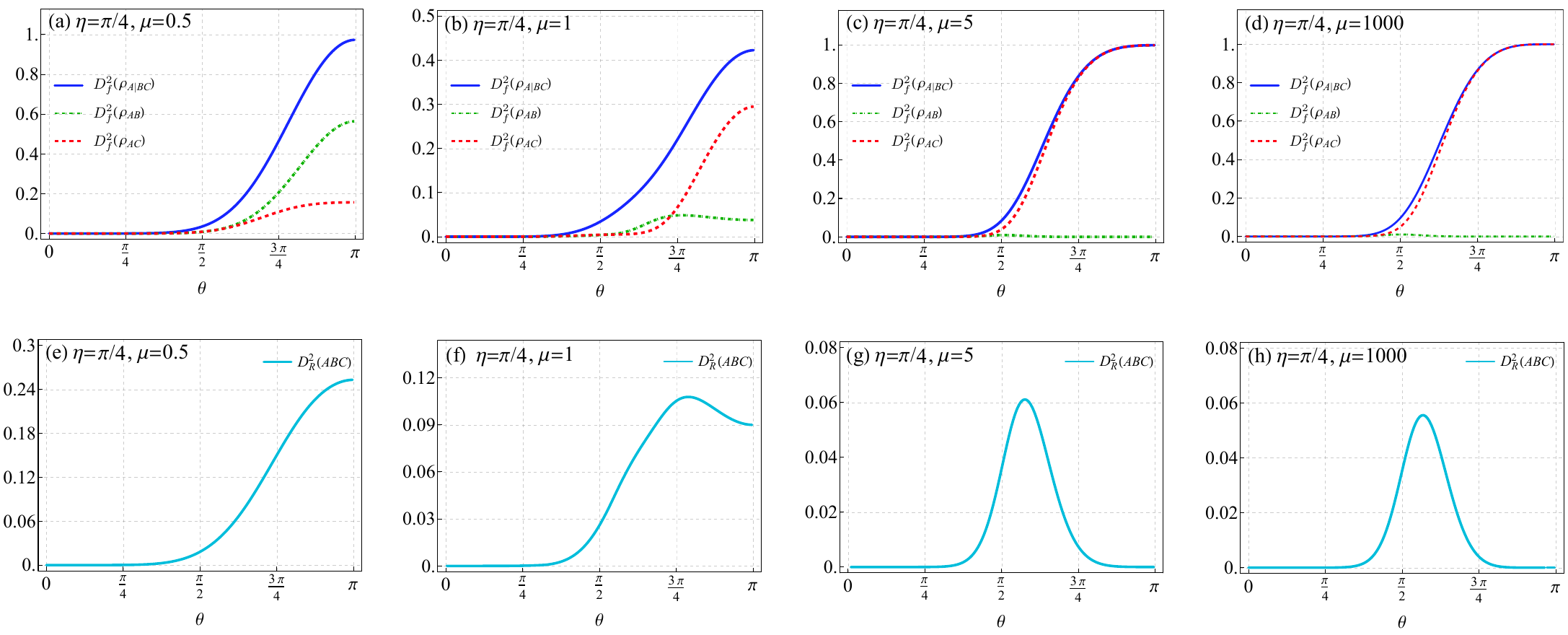}
		\caption{Monogamy relation of quantum discord is depicted as a function of scattering angle $\theta$ for different values of the scattering momentum $\mu$ at $\eta=\pi/4$. 
			Herein, the solid blue curve in panels (a)–(d) represents $ D_{f}^{2}(\rho_{A|BC})$, the green dash-dotted curve represents $ D_{f}^{2}(\rho_{AB})$, the red dashed curve represents $ D_{f}^{2}(\rho_{AC})$.
			Panels (e)–(h) plot the  $D_R^{2}\left(\rho_{ABC}\right)$ as a function of $\theta$ for different values of  $\mu$.}
		\label{Fig7}
	\end{figure*}
	where this residual quantity corresponds to the difference between the global bipartite entanglement of $A|BC$ and the sum of the local bipartite entanglements of $AB$ and $AC$. It quantifies the component of tripartite entanglement in pure states that cannot be accounted for by bipartite entanglement contributions alone.
	
	Fig. \ref{Fig5} illustrates the monogamy relations and their evolutionary characteristics as quantified by the SEF. 
	Figs. \ref{Fig5}\textcolor{blue}{(a)-(d)} depict the three bipartite entanglement quantities, $E_{f}^{2}(\rho_{A|BC})$, $E_{f}^{2}(\rho_{AB})$, and $E_{f}^{2}(\rho_{AC})$, as functions of the scattering angle $\theta$ and the scattering momentum $\mu$ for $\eta=\pi/4$. Our findings reveal that these three quantities exhibit distinctly different evolutionary trends: 
	$E_{f}^{2}(\rho_{A|BC})$ varies non-monotonically with the scattering momentum $\mu$ and eventually plateaus at a saturation value in the relativistic limit; meanwhile, $E_{f}^{2}(\rho_{AB})$ and $E_{f}^{2}(\rho_{AC})$ display a clear competitive behavior, where the former asymptotically vanishes with increasing $\mu$ while the latter gradually rises and eventually converges with the global entanglement $E_{f}^{2}(\rho_{A|BC})$. 
	Furthermore, Figs. \ref{Fig5}\textcolor{blue}{(e)-(h)} confirm that despite the diverse evolution of each bipartite term, the residual entanglement consistently satisfies the non-negativity condition $E_{R}^{2}(\rho_{ABC}) \ge 0$, strictly adhering to the fundamental constraints of quantum monogamy. 
	
	This correlation competition induced by monogamy intuitively highlights the physical advantages of the concurrence fill metric. This measure interprets the three bipartite entanglements as the side lengths of an $entanglement$ $triangle$, whose area reaches its maximum when the side lengths are equal. As previously noted, an increase in one bipartite correlation (e.g., $E_f(\rho_{AC})$) inevitably leads to the depletion of another (e.g., $E_f(\rho_{AB})$) due to monogamy constraints. This extreme imbalance causes the entanglement triangle to collapse, significantly suppressing the residual entanglement and thereby explaining the corresponding reduction in the global tripartite entanglement of the system.
	It is essential to emphasize that within this standard quantum mechanical framework, the non-negativity of the residual entanglement is mathematically guaranteed. Therefore, its physical significance lies not in the mere satisfaction of the inequality, but in the magnitude of the residual value $E_{R}^{2}(\rho_{ABC})$ itself. A larger residual value indicates a relaxation of monogamy constraints, allowing quantum resources to be more evenly distributed and shared across the global $ABC$ system. Conversely, a smaller residual value indicates that entanglement is tightly locked within specific bipartite pairs, fundamentally restricting multi-particle collaborative shareability. As our results demonstrate, such high shareability is primarily achieved within the non-relativistic regime.

	Fig. \ref{Fig6} depicts the variation of the residual entanglement  $E_{R}^{2}\left(\rho_{ABC}\right)$  with the scattering angle $\theta$ for different values of $\eta$ and $\mu$. Fig. \ref{Fig6}\textcolor{blue}{(a)} reveals that for fixed $\mu$, $ E_{R}^{2}\left(\rho_{ABC}\right) $ increases with rising $\eta$. As seen in Fig. \ref {Fig6}\textcolor {blue}{(b)}, the residual entanglement exhibits monotonic decay as the scattering momentum $\mu$ increases for a fixed $\eta$.
	
	\subsection{Monogamy relation for SQD}\label{IIIC}	
	We next investigate the monogamy of quantum discord in this scattering model. Analogous to entanglement monogamy, we define the residual discord as
	\begin{equation}
		D_{R}^{2}\left(\rho_{A B C}\right)=D^{2}_f\left(\rho_{A \mid B C}\right)-D^{2}_f\left(\rho_{A B}\right)-D^{2}_f\left(\rho_{A C}\right).
	\end{equation}		
	Fig. \ref {Fig7} presents the monogamy relation for the SQD as a function of the scattering angle $\theta$.
	Figs. \ref{Fig7}\textcolor{blue}{(a)–(d)} depict the evolution of three bipartite quantum discords quantified by the SQD, namely  $D^{2}_f\left(\rho_{A|B C}\right)$, $D^{2}_f\left(\rho_{AB }\right)$, and $D^{2}_f\left(\rho_{AC}\right)$, as functions of the scattering angle $\theta$ and the scattering momentum $\mu$ for  $\eta=\pi/4$. These quantities exhibit evolution trends analogous to those of the SEF.
	Figs. \ref{Fig7}\textcolor{blue}{(e)–(h)} illustrate the dependence of the residual discord $ D_{R}^{2}\left(\rho_{A B C}\right)$ versus the scattering angle $\theta$ for various values of the scattering momentum $\mu$.
	We find that $ D_{R}^{2}\left(\rho_{A B C}\right)\ge0$ holds universally, which confirms that the monogamy relation Eq. (\ref{Eq.2.2.20}) is always satisfied in this scattering model. Additionally, in the non-relativistic regime, the monogamy relations for SQD become more relaxed, leading to a larger residual discord and consequently higher shareability of quantum discord in the global system.
	Quantum discord encompasses all non-classical correlations, including the strong correlations of entangled states and the weak quantum correlations of separable states, whereas quantum entanglement only characterizes the strong quantum correlations of inseparable states. Quantum entanglement thus constitutes a subset of quantum discord, for the same quantum state, the following inequality holds
	\begin{equation}
		D_{R}^{2}\left(\rho_{A B C}\right)\ge E_{R}^{2}\left(\rho_{A B C}\right).
	\end{equation} 
	A direct comparison between Figs. \ref{Fig7}\textcolor{blue}{(e)–(h)} and  Figs. \ref{Fig5}\textcolor{blue}{(e)–(h)}  confirms that residual  discord is consistently larger than residual  entanglement, the validity of the above inequality has been verified.
	\begin{figure}[htbp]
		\centering
		\includegraphics[width=7cm]{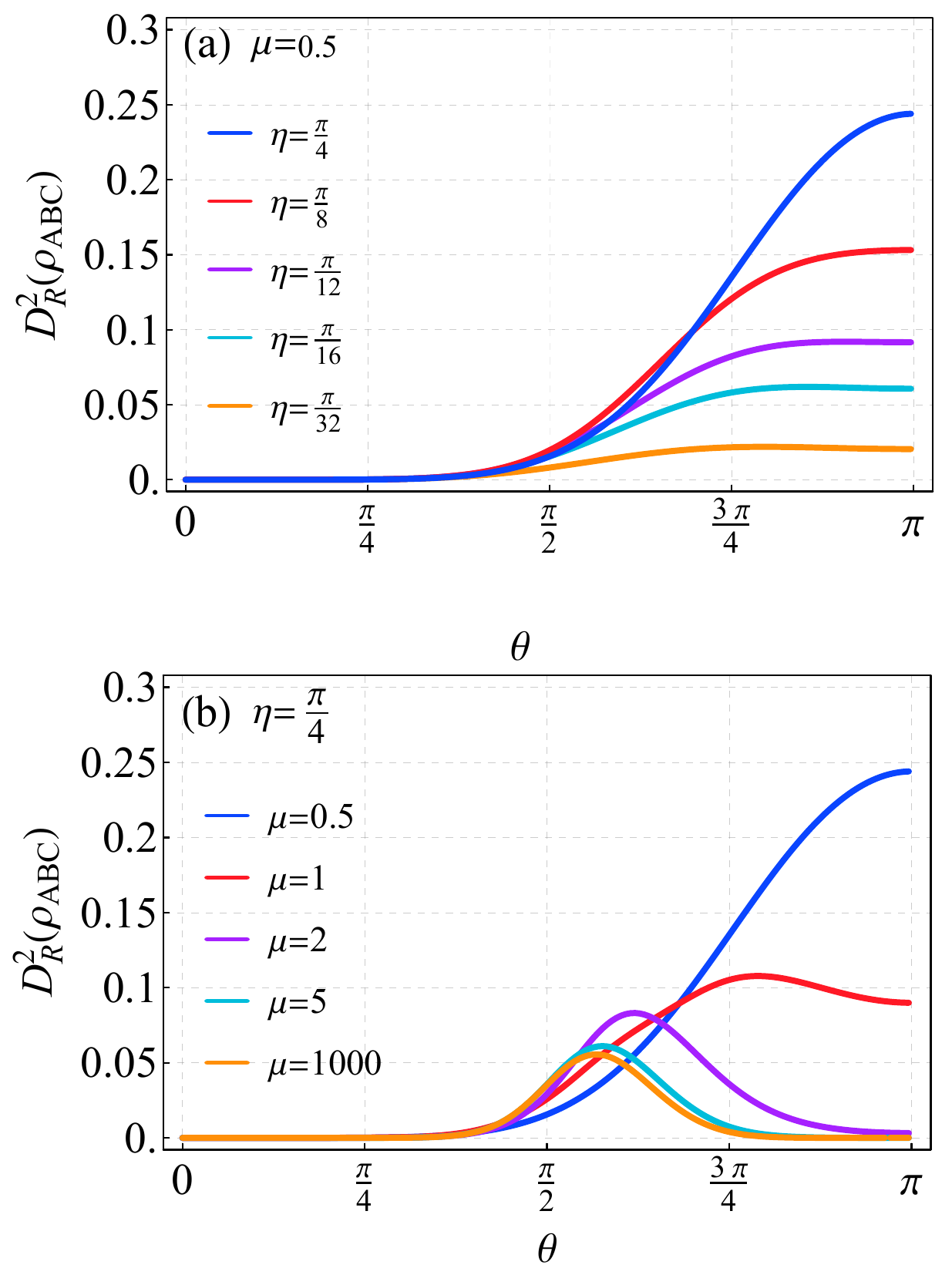}
		\caption{Residual discord is plotted as a function of scattering angle $\theta$ for specific scattering scattering momentum $\mu$ and initial entanglement weight $\eta$.}
		\label{Fig8}
	\end{figure}
	
	Fig. \ref{Fig8} illustrates the dependence of the residual discord on the scattering angle $\theta$ for various values of  $\mu$ and $\eta$. 
	In Fig. \ref{Fig8}\textcolor{blue}{(a)}, for fixed $\mu=0.5$, $D_{R}^{2}\left(\rho_{ABC}\right)$ increases monotonically with $\eta$. 
	In Fig. \ref{Fig8}\textcolor{blue}{(b)}, for fixed $\eta=\pi/4$, the residual discord $D_{R}^{2}\left(\rho_{ABC}\right)$ decays monotonically with increasing scattering momentum $\mu$.
	Therefore, the strong $B$-$C$ correlation and the relativistic limit suppress the generation of residual discord and reduce the shareability of quantum discord in the global system.
	
	\section{CONCLUSIONS AND DISCUSSIONS} \label{IV}
	In this work, we systematically investigate the generation of GTE and the monogamy properties of quantum correlations in tree-level Bhabha scattering with an entangled spectator particle, a configuration that has previously been studied only in the context of bipartite entanglement transfer, with the dynamics of global multipartite entanglement largely unexplored. 
	
	Our core physical findings are summarized as follows: First, we identify two indispensable resources for GTE generation: non-zero initial $B$-$C$ subsystem entanglement (parameterized by $\eta$) and scattering momentum $\mu$ of the electron-positron pair. GTE is strictly zero in the trivial limits of $\eta=0$ ($B$-$C$ product state) or $\mu=0$ (strict nonrelativistic limit), where spin-dependent correlations in the Bhabha scattering amplitude vanish entirely, eliminating the entanglement transfer mechanism. For finite $\eta$ and $\mu$, GTE varies non-monotonically with both parameters, peaking at intermediate values ($\theta_{\max}\approx2.21$, $\eta_{\max}\approx1.42$, $\mu_{\max}\approx0.913$) rather than at the maximally entangled $B$-$C$ initial state or the ultrarelativistic limit. All four canonical tripartite entanglement measures yield fully consistent GTE evolution, and concurrence fill offers unique advantages by integrating global entanglement and pairwise correlation distributions, encoding more comprehensive quantum information than extremal-value-based measures. We further confirm all results for GTE, entanglement of formation and quantum discord are strictly independent of the initial phase factor $\beta$, demonstrating the robustness of our findings. Second, we characterize quantum correlation monogamy via residual entanglement and residual discord. Rather than focusing on the mathematically trivial validity of monogamy inequalities for pure quantum states, we center our analysis on the physical meaning of these residuals and their intrinsic link to GTE. Residual entanglement/discord quantifies the global tripartite correlation that cannot be explained by pairwise bipartite correlations: a larger residual value signals relaxed monogamy constraints and more even correlation distribution across the three particles, while a smaller value indicates tightened constraints, with correlations locked in a single bipartite pair and global tripartite correlations strongly suppressed. Monogamy constraint strength correlates strongly with scattering momentum: constraints are markedly relaxed in the non-relativistic regime to enhance correlation shareability, and drastically tightened in the ultrarelativistic regime where both residual correlations and GTE are suppressed. Critically, residual entanglement magnitude has a strict positive correlation with GTE strength across all parameter space, and non-zero residual entanglement (relaxed monogamy) is a necessary prerequisite for GTE generation, providing an intuitive physical explanation for the non-monotonic GTE evolution. We also confirm residual discord is consistently larger than residual entanglement across all configurations, consistent with quantum discord capturing all non-classical correlations beyond entanglement alone.
	
	This work has direct experimental relevance for both current and future platforms, supported by recent breakthroughs in quantum correlation measurements in high-energy collisions. For instance, recent analyses by the ATLAS and CMS collaborations at the CERN LHC have unambiguously observed quantum entanglement between top-antitop quark pairs in proton-proton collisions \cite{2024ATLAS,2024CMS1,2024CMS2}. These measurements verified spin correlation strengths exceeding the classical limit of local hidden variable theories, marking the first definitive confirmation of quantum entanglement in TeV-scale scattering, and proving that quantum correlation signals can be reliably extracted from collider event data. This lays a critical experimental foundation for the observation of GTE in Bhabha scattering proposed in this work. For next-generation high-luminosity lepton colliders including the CEPC and FCC-ee \cite{CEPC2025,FCC-ee2019}, Bhabha scattering is the primary process for accelerator luminosity calibration with extremely high event rates, making these facilities ideal platforms to test our tripartite entanglement predictions.
	From the perspective of quantum information applications, this model is a direct high-energy analog of realistic remote entanglement distribution and entanglement swapping protocols for quantum networks \cite{Kimble2008,Zhong2021}. In these practical protocols, one particle of a pre-entangled pair is stored in a quantum memory as the stationary spectator (corresponding to particle $C$ in our model), while its entangled partner (particle $B$) is transmitted through a quantum channel and undergoes controlled coherent scattering with an incident particle (particle $A$). This is exactly the core physical configuration we study in this work,  meaning our results may provide theoretical guidance for the parameter design of such protocols.	    
	There are several impactful extensions of this work for future research. Our current analysis is based on the tree-level QED approximation, so a critical next step is to investigate the impact of higher-order loop, radiative and inelastic scattering corrections on GTE evolution, to assess the robustness of our predictions. It is also promising to extend this framework to other fundamental QED processes such as Møller and Compton scattering, to explore the universality of tripartite entanglement generation in electromagnetic interactions.

	Overall, this work uncovers novel multipartite quantum correlation properties inherent to fundamental QED scattering processes, and establishes a clear, experimentally testable link between high-energy particle physics and quantum information science. Our findings not only advance the theoretical understanding of quantum correlations in relativistic scattering, but also provide a new framework for exploring quantum resources in high-energy experiments and developing quantum information protocols based on fundamental particle interactions.

	\begin{acknowledgements}	
    This work was supported by the National Natural Science Foundation of China (Grants No. 12475009, and No. 12075001),  Anhui Province Science and Technology Innovation Project (Grant No. 202423r06050004), Anhui Provincial Natural Science Foundation (Grant No.~ 2508085ZD001), and Anhui Provincial University Scientific Research Major Project (Grant No. 2024AH040008), and Anhui Provincial Department of Industry and Information Technology (Grant No. JB24044).  
	\end{acknowledgements}
	
	\appendix
	\begin{widetext}
		\section{Bhabha scattering amplitudes}\label{A}
		{The Bhabha scattering amplitude is calculated in the CM reference frame of particles $A$ and $B$ denote as  
			\begin{equation}
				\mathcal{M}_{\mathrm{Bhabha}}=e^2\left(\bar{v}(b,p_2)\gamma^\mu u(a,p_1)\frac{1}{(p_1+p_2)^2}\bar{u}(r,p_3)\gamma_\mu v(s,p_4)-\bar{v}(b,p_2)\gamma^\mu v(s,p_4)\frac{1}{(p_3-p_1)^2}\bar{u}(r,p_3)\gamma_\mu u(a,p_1)\right),
			\end{equation}
			where $a, b, r, s$ are the spin indices. According to  Eqs. (\ref{Eq.55})-(\ref{Eq.66}) the explicit expressions for the polarized amplitudes are given by 		\begin{equation}\mathcal{M}(RR;RR)=\mathcal{M}(LL;LL)=\frac{(2+11\mu^2+8\mu^4+2\cos\theta+\mu^2\cos2\theta)\csc^2(\frac{\theta}{2})}{4\mu^2(1+\mu^2)},\end{equation}
			\begin{equation}\mathcal{M}(RR;_{LR}^{RL})=-\mathcal{M}(LL;_{LR}^{RL})=-\frac{(1+\mu^2\cos\theta)\cot(\frac{\theta}{2})}{\mu^2\sqrt{1+\mu^2}},\end{equation}
			\begin{equation}\mathcal{M}(RR;LL)=\mathcal{M}(LL;RR)=\frac{1+\mu^2(1+\cos\theta)}{\mu^2(1+\mu^2)},\end{equation}
			\begin{equation}\mathcal{M}(_{LR}^{RL};RR)=-\mathcal{M}(_{LR}^{RL};LL)=\frac{(1+\mu^2\cos\theta)\cot(\frac{\theta}{2})}{\mu^2\sqrt{1+\mu^2}},\end{equation}
			\begin{equation}\mathcal{M}(RL;RL)=\mathcal{M}(LR;LR)=\frac{(1+\mu^2(1+\cos\theta))\cot^2(\frac{\theta}{2})}{\mu^2},\end{equation}
			\begin{equation}\mathcal{M}(RL;LR)=\mathcal{M}(LR;RL)=1-\cos\theta-\frac{1}{\mu^2}.\end{equation}}
		
		\section{Reduce density matrix}\label{B}
		From Eqs. (\ref{Eq.2.1.2})-(\ref{Eq.2.1.5}), we can compute the various reduced density matrices for the initial and final states
		\begin{equation}
			\rho_{AB}^{i}=\cos^2\eta|R\rangle_A|R\rangle_{BB}\langle R|_A\langle R|+\sin^2\eta|R\rangle_A|L\rangle_{BB}\langle L|_A\langle R|,
		\end{equation}
		\begin{equation}
			\rho_{AC}^{i}=\cos^2\eta|R\rangle_A|R\rangle_{CC}\langle R|_A\langle R|+\sin^2\eta|R\rangle_A|L\rangle_{CC}\langle L|_A\langle R|, 
		\end{equation}
		\begin{equation}\begin{aligned}
				& \rho_{BC}^{i}=\mathrm{cos}^{2}\eta|R\rangle_{B}|R\rangle_{CC}\langle R|_{B}\langle R|+e^{-i\beta}\sin\eta\operatorname{cos}\eta|R\rangle_{B}|R\rangle_{CC}\langle L|_{B}\langle L| \\
				& +e^{i\beta}\sin\eta\cos\eta|L\rangle_B|L\rangle_{CC}\langle R|_B\langle R|+\mathrm{sin}^2\eta|L\rangle_B|L\rangle_{CC}\langle L|_B\langle L|,
			\end{aligned}
		\end{equation}
		\begin{equation}\begin{aligned}
				\rho_{AB}^{f}&=\sum_{r,s,r^{\prime},s^{\prime}}\left[\cos^2\eta\mathcal{M}(RR;rs)\mathcal{M}^{\dagger}(RR;r^{\prime}s^{\prime})|r\rangle_A|s\rangle_{BB}\langle s^{\prime}|_A\langle r^{\prime}|\right.\\
				&\left.+\sin^2\eta\mathcal{M}(RL;rs)\mathcal{M}^\dagger(RL;r^{\prime}s^{\prime})|r\rangle_A|s\rangle_{BB}\langle s^{\prime}|_A\langle r^{\prime}|\right],
			\end{aligned}
		\end{equation}
		\begin{equation}\begin{aligned}
				&\rho_{AC}^{f}=\sum_{r,r^{\prime},s}\Bigl[\mathrm{cos}^{2}\eta\mathcal{M}(RR;rs)\mathcal{M}^{\dagger}(RR;r^{\prime}s)|r\rangle_{A}|R\rangle_{CC}\langle R|_{A}\langle r^{\prime}| \\
				& +e^{-i\beta}\sin\eta\cos\eta\mathcal{M}(RR;rs)\mathcal{M}^\dagger(RL;r's)|r\rangle_A|R\rangle_{CC}\langle L|_A\langle r'| \\
				& +e^{i\beta}\sin\eta\cos\eta\mathcal{M}(RL;rs)\mathcal{M}^\dagger(RR;r^{\prime}s)|r\rangle_A|L\rangle_{CC}\langle R|_A\langle r^{\prime}| \\
				& +\sin^2\eta\mathcal{M}(RL;rs)\mathcal{M}^\dagger(RL;r^\prime s)|r\rangle_A|L\rangle_{CC}\langle L|_A\langle r^\prime|\Big],
			\end{aligned}
		\end{equation}
		\begin{equation}\begin{aligned}
				\rho_{BC}^{f} & =\sum_{r,r,s^{\prime}}\bigl[\mathrm{cos}^{2}\eta\mathcal{M}(RR;rs)\mathcal{M}^{\dagger}(RR;rs^{\prime})|s\rangle_{B}|R\rangle_{CC}\langle R|_{B}\langle s^{\prime}| \\
				& +e^{-i\beta}\sin\eta\cos\eta\mathcal{M}(RR;rs)\mathcal{M}^\dagger(RL;rs^\prime)|s\rangle_B|R\rangle_{CC}\langle L|_B\langle s^\prime| \\
				& +e^{i\beta}\sin\eta\cos\eta\mathcal{M}(RL;rs)\mathcal{M}^\dagger(RR;rs^{\prime})|s\rangle_B|L\rangle_{CC}\langle R|_B\langle s^{\prime}| \\
				& +\sin^2\eta\mathcal{M}(RL;rs)\mathcal{M}^\dagger(RL;rs^\prime)|s\rangle_B|L\rangle_{CC}\langle L|_B\langle s^\prime|\Big].
			\end{aligned}
		\end{equation}
	\end{widetext}

	\newpage
	\appendix 

\begin{references} 
		\bibitem{1.1QM1935}
		A. Einstein, B. Podolsky, and N. Rosen,
		\href{https://link.aps.org/doi/10.1103/PhysRev.47.777}
		{Phys. Rev. {\bf 47}, 777  (1935)}.
		\bibitem{1.2QM1935}
		E. Schrodinger,
		\href{https://www.cambridge.org/core/journals/mathematical-proceedings-of-the-cambridge-philosophical-society/article/discussion-of-probability-relations-between-separated-systems/C1C71E1AA5BA56EBE6588AAACB9A222D}
		{Proc. Cambridge Philos. Soc. {\bf 31}, 555  (1935)}.    
		\bibitem{1.7RevModPhys.81.865}
		R. Horodecki, P. Horodecki, M. Horodecki, and K. Horodecki, 
		\href{https://link.aps.org/doi/10.1103/RevModPhys.81.865}
		{Rev. Mod. Phys. {\bf  81}, 865  (2009)}.
		
		\bibitem{1.5Pasquale2004}
		P. Calabrese and J. Cardy,
		\href{https://doi.org/10.1088/1742-5468/2004/06/P06002}
		{J. Stat. Mech. {\bf  8}, P06002  (2004)}.
		\bibitem{1.6Casini2009}
		H. Casini and M. Huerta
		\href{https://doi.org/10.1088/1751-8113/42/50/504007}
		{J. Phys. A {\bf 49}, 504007 (2009)}.
		\bibitem{1.8Book}
		D. R. Terno, \textit{Quantum Information Processing: From Theory to Experiment}, edited by D. G. Angelakis et al.(IOP Press, Amsterdam, 2006).
		\href{}
		{{\bf }}
		\bibitem{1.9PhysRevLett.122.102001}
		S. R. Beane, D. B. Kaplan, N. Klco, and M. J. Savage,
		\href{https://link.aps.org/doi/10.1103/PhysRevLett.122.102001}
		{Phys. Rev. Lett.  {\bf  112},  102001 (2019)}.
		\bibitem{1.10PhysRevD.97.016011}
		J. Fan and X. Li,
		\href{https://link.aps.org/doi/10.1103/PhysRevD.97.016011}
		{Phys. Rev. D  {\bf  97},  016011 (2018)}.
		
		
		\bibitem{1.11JHEP}
		R. Aoude, E. Madge, F. Maltoni, and L. Mantani,
		\href{https://doi.org/10.1007/JHEP12(2023)017}
		{J. High Energy Phys.   {\bf  12},  017  (2023)}.
		\bibitem{1.12PhysRevD.108.025015}
		A. Sinha and A. Zahed,
		\href{https://link.aps.org/doi/10.1103/PhysRevD.108.025015}
		{Phys. Rev. D  {\bf  108},  025015 (2023)}.
		\bibitem{1.13PhysRevC.108.L041601}
		G. A. Miller, 
		\href{https://link.aps.org/doi/10.1103/PhysRevC.108.L041601}
		{Phys. Rev. C  {\bf  108},  L041601  (2023)}.
		\bibitem{1.14PhysRevD.104.116021}
		J. Fan, G. M. Deng, and X. J. Ren,
		\href{https://link.aps.org/doi/10.1103/PhysRevD.104.116021}
		{Phys. Rev. D  {\bf  104},  116021  (2021)}.
		\bibitem{1.15PESCHANSKI201689}
		R. Peschanski and S. Seki,
		\href{https://www.sciencedirect.com/science/article/pii/S0370269316301423}
		{Phys. Lett. B   {\bf  758},  89   (2016)}.
		\bibitem{1.16PhysRevD.107.116007}
		S. Fedida and A. Serafini, 
		\href{https://link.aps.org/doi/10.1103/PhysRevD.107.116007}
		{Phys. Rev. D   {\bf  107},   116007 (2023)}.
		\bibitem{1.17PhysRevLett.125.181602}
		R. Aoude, M. Z. Chung, Y. T. Huang, C. S. Machado, and M. K. Tam, 
		\href{https://link.aps.org/doi/10.1103/PhysRevLett.125.181602}
		{Phys. Rev. Lett.   {\bf  125}, 181602   (2020)}.
		\bibitem{1.18PhysRevD.95.114008}
		D. E. Kharzeev and E. M. Levin,
		\href{https://link.aps.org/doi/10.1103/PhysRevD.95.114008}
		{Phys. Rev. D   {\bf  95},   114008   (2017)}.
		\bibitem{3.7TreelevelQRDPhysRevD2025}
		M. Blasone, S.  De Siena, G.  Lambiase, C. Matrella,  and B. Micciola, 
		\href{https://link.aps.org/doi/10.1103/PhysRevD.111.016007}
		{Phys. Rev. D   {\bf 111},  016007 (2025)}.
		\bibitem{3.8QuarkFisher2024}	
		B. L. Ye,  L. Y. Xue, Z. Q. Zhu, D. D.  Shi, and S. M. Fei, 
		\href{https://link.aps.org/doi/10.1103/PhysRevD.110.055025}
		{Phys. Rev. D  {\bf 110},  055025 (2024)}.
		\bibitem{3.9QuarkBell2021}
		M. Fabbrichesi,  R. Floreanini, and G. Panizzo, 
		\href{https://link.aps.org/doi/10.1103/PhysRevLett.127.161801}
		{Phys. Rev. Lett.  {\bf  127}, 161801  (2021)}.
		\bibitem{3.10Bernabeu2015}
		J. Bernabeu, A. Di Domenico, and P. Villanueva-Perez,
		\href{https://doi.org/10.1007/JHEP10(2015)139}
		{J. High Energy Phys.  {\bf }10  (2015) 139}.
		\bibitem{3.11Lello}
		L. Lello, D. Boyanovsky, and R. Holman, 
		\href{https://doi.org/10.1007/JHEP11(2013)116}
		{J. High Energy Phys.  {\bf }11  (2013) 116}.
		
		
		\bibitem{2.7Afik2022quantuminformation}
		Y. Afik and J. R. Muñoz de Nova
		\href{https://doi.org/10.22331/q-2022-09-29-820}
		{ Quantum  {\bf  16},  820     (2022)}.
		\bibitem{2.8PhysRevLett.130.221801}
		Y. Afik and J. R. Muñoz de Nova 
		\href{https://link.aps.org/doi/10.1103/PhysRevLett.130.221801}
		{Phys. Rev. Lett.  {\bf  130}, 221801     (2023)}.
		\bibitem{2.9fhkc-kfhr}
		Y. Afik,  Y. Kats, J. R. Muñoz de Nova, A. Soffer, and  D. Uzan,
		\href{https://link.aps.org/doi/10.1103/fhkc-kfhr}
		{Phys. Rev. D {\bf  111}, L111902 (2025)}.
	
	
		\bibitem{2.1Blasone20141}
		M. Blasone, F. Dell'Anno, S. De Siena, and F. Illuminati, 
		\href{https://dx.doi.org/10.1209/0295-5075/106/30002}
		{Europhys. Lett.  {\bf  106},   30002   (2014)}.
		\bibitem{2.2Blasone20142}
		M. Blasone,  F. Dell'Anno, S.  De Siena,  and F. Illuminati, 
		\href{https://onlinelibrary.wiley.com/doi/abs/10.1155/2014/359168}
		{Adv. High Energy Phys. {\bf  2014},  359168  (2014)}.
		\bibitem{2.3Bittencourt2023}
		V. Bittencourt, M.  Blasone, and G. Zanfardino, 
		\href{https://dx.doi.org/10.1088/1742-6596/2533/1/012024}
		{J. Phys. Conf. Ser.    {\bf  2533},  012024    (2023)}.
		\bibitem{2.4PhysRevLett.117.050402}
		J. A. Formaggio, D. I.  Kaiser, M. M. Murskyj, and T. E. Weiss,
		\href{https://link.aps.org/doi/10.1103/PhysRevLett.117.050402}
		{Phys. Rev. Lett.   {\bf  117},   050402   (2016)}.
		\bibitem{2.5PhysRevD.99.095001}
		J. Naikoo, A. K. Alok, S.  Banerjee, and S. U. Sankar, 
		\href{https://link.aps.org/doi/10.1103/PhysRevD.99.095001}
		{Phys. Rev. D  {\bf 99}, 095001     (2019)}.
		\bibitem{2.6PhysRevA.108.032210}
		M. Blasone, F.  Illuminati, L.  Petruzziello,  and L. Smaldone, 
		\href{https://link.aps.org/doi/10.1103/PhysRevA.108.032210}
		{Phys. Rev. A  {\bf  108},  032210    (2023)}.
		\bibitem{2.6.1PhysRevA.88.022115}
		D. Gangopadhyay, D. Home, and A. S. Roy,
		\href{https://link.aps.org/doi/10.1103/PhysRevA.88.022115}
		{Phys. Rev. A    {\bf   88},   022115   (2013)}.
		\bibitem{2.6.2PhysRevA.98.050302}
		X. K. Song, Y. Huang, J. Ling, and M. H. Yung, 
		\href{https://link.aps.org/doi/10.1103/PhysRevA.98.050302}
		{Phys. Rev. A   {\bf  98},   050302 (2018)}.
		\bibitem{2.6.3MingFei}
		F. Ming, X. K. Song, J. Ling, L. Ye, and D. Wang,
		\href{https://doi.org/10.1140/epjc/s10052-020-7840-y}
		{Eur. Phys. J. C   {\bf  80},   275   (2020)}.
		\bibitem{2.6.4WangDong}
		D. Wang, F. Ming, X. K. Song, L. Ye, and J. L. Chen, 
		\href{https://doi.org/10.1140/epjc/s10052-020-8403-y}
		{Eur. Phys. J. C  {\bf   800},   800 (2020)}.
		\bibitem{2.6.5Liyuwen}
		Y. W. Li, L. J. Li, X. K. Song,  D. Wang, and L. Ye, 
		\href{https://doi.org/10.1140/epjc/s10052-022-10759-2}
		{Eur. Phys. J. C  {\bf  82}, 799 (2022)}.
		\bibitem{2.6.6Wangguangjie}
		G. J. Wang, L. J. Li, X. K. Song, and  D. Wang,
		\href{https://doi.org/10.1140/epjc/s10052-023-11979-w}
		{Eur. Phys. J. C  {\bf  83},  801 (2023)}.


		\bibitem{2.9SciPostPhys.3.5.036}
		A. Cervera-Lierta, J. I. Latorre, J. Rojo, and L. Rottoli,
		\href{https://scipost.org/10.21468/SciPostPhys.3.5.036}
		{SciPost Phys.  {\bf 3},   036   (2017)}.
		\bibitem{2.10Cervera2019}
		A. Cervera-Lierta,
		\href{https://arxiv.org/abs/1906.12099}
		{arXiv:1906.12099{\bf}}.
		\bibitem{2.11Entanglementsaturation2025}
		B. Massimo, G. L. Silvio De Siena,  M. Cristina, and M. Bruno,
		\href{https://doi.org/10.48550/arXiv.2505.06878}
		{arXiv:2505.06878{\bf}}.
		
		
		\bibitem{2.12spectatorPhysRevD.2019}
		J. B. Araujo, B. Hiller, I. G. da Paz, Manoel M. Ferreira, Marcos Sampaio, and H. A. S. Costa, 
		\href{https://link.aps.org/doi/10.1103/PhysRevD.100.105018}
		{Phys. Rev. D  {\bf  100}, 105018  (2019)}.
	    spectator particle
		\bibitem{2.13spectatorPhysRevD.2022}
		J. D. Fonseca, B. Hiller, J. B. Araujo, I. G. da Paz, and M.
		Sampaio,
		\href{https://link.aps.org/doi/10.1103/PhysRevD.106.056015}
		{Phys. Rev. D  {\bf  106},   056015  (2022)}.
		\bibitem{2.14spectatorPhysRevD2024}
		M. Blasone, G. Lambiase, and B. Micciola,
		\href{https://link.aps.org/doi/10.1103/PhysRevD.109.096022}
		{Phys. Rev. D  {\bf  109},   096022  (2024)}.
		
		\bibitem{3.1RevModPhys.91.025001}
		E. Chitambar and G. Gour,
		\href{https://link.aps.org/doi/10.1103/RevModPhys.91.025001}
		{Rev. Mod. Phys.  {\bf 91},  025001 (2019)}.
	
		\bibitem{4.1GGMPhysRevA.81.012308}
		A. Sen(De) and U. Sen, 
		\href{https://link.aps.org/doi/10.1103/PhysRevA.81.012308}
		{Phys. Rev. A  {\bf  81},  012308  (2010)}.
		\bibitem{4.2GGMPhysRevA.94.022336}
		T. Das, S. Singha Roy, S. Bagchi, A. Misra, A. Sen(De), and U. Sen,
		\href{https://link.aps.org/doi/10.1103/PhysRevA.94.022336}
		{Phys. Rev. A  {\bf  94 }, 022336 (2016)}.
		\bibitem{4.3GGMPhysRevA.95.022301}
		D. Sadhukhan, S. S. Roy, A. K. Pal, D. Rakshit, A. Sen(De), and U. Sen,
		\href{https://link.aps.org/doi/10.1103/PhysRevA.95.022301}
		{Phys. Rev. A  {\bf  95},  022301 (2017)}.
		
		
		\bibitem{4.4Threepi2007}
		Y. C. Ou and  H. Fan,
		\href{http://dx.doi.org/10.1103/PhysRevA.75.062308}
		{Phys. Rev. A  {\bf  75},  062308 (2007)}.
		
		\bibitem{4.5GMC2011}
		Z. H. Ma, Z. H. Chen, and J. L. Chen,
		\href{https://link.aps.org/doi/10.1103/PhysRevA.83.062325}
		{Phys. Rev. A  {\bf  83},   062325   (2011)}.
		
		\bibitem{4.6Concurrencefill}
		S. B. Xie and J. H. Eberly, 
		\href{https://link.aps.org/doi/10.1103/PhysRevLett.127.040403}
		{Phys. Rev. Lett. {\bf  127},  040403  (2021)}.
		
		\bibitem{4.8SEFPhysRevLett.113.100503}
		Y. K. Bai, Y. F. Xu, and Z. D. Wang,
		\href{https://link.aps.org/doi/10.1103/PhysRevLett.113.100503}
		{Phys. Rev. Lett.  {\bf  113},  100503  (2014)}.
	
		\bibitem{4.9SQDPhysRevA.88.012123}
		Y. K. Bai, N. Zhang, M. Y. Ye, and Z. D. Wang, 
		\href{https://link.aps.org/doi/10.1103/PhysRevA.88.012123}
		{Phys. Rev. A  {\bf  88},  012123  (2013)}.
	
		\bibitem{EntanglementSwapping}
		S. M. Zangi, C. Shukla, A. u. Rahman, and  B. Zheng,
		\href{https://doi.org/10.3390/e25030415}
		{Entropy {\bf  25}, 415 (2023)}.
		\bibitem{Remote4x8dcmyx}
		T. L. Wang, P. Wang, Z. A. Zhao, S. Zhang, R. Z. Zhao, X. Y. Yang, H. F. Zhang, Z. F. Li, and Y. Wu,
		\href{https://link.aps.org/doi/10.1103/4x8d-cmyx}
		{PRX Quantum {\bf 7}, 010348 (2026)}.
		
		
		
		
		
		
		
		
		
		
		
		
		
		
		
		
		
		
		\bibitem{5.1GMIPhysRevLett.96.220503}
		T. J. Osborne and F. Verstraete, 
		\href{https://link.aps.org/doi/10.1103/PhysRevLett.96.220503}
		{Phys. Rev. Lett.   {\bf  96}, 20503 (2006)}.
		
		\bibitem{5.2Globalentanglement}
		D. A. Meyer and N. R. Wallach,
		\href{https://doi.org/10.1063/1.1497700}
		{J. Math. Phys. {\bf  43}, 4273 (2002)}.
		\bibitem{5.3Anobservablemeasure}
		G. K. Brennen,
		\href{https://doi.org/10.26421/QIC3.6-5}
		{Quantum Inf. Comput. {\bf  3}, 619 (2003)}.
		\bibitem{5.4PhysRevA.61.052306}
		V. Coffman, J. Kundu, and W. K. Wootters,
		\href{https://link.aps.org/doi/10.1103/PhysRevA.61.052306}
		{Phys. Rev. A  {\bf  61}, 052306(2000)}.
		\bibitem{6.1PhysRevLett.80.2245}
		W. K. Wootters,
		\href{https://link.aps.org/doi/10.1103/PhysRevLett.80.2245}
		{Phys. Rev. Lett. {\bf  88}, 2245 (1998)}.
		\bibitem{6.2PhysRevLett.88.017901}
		H. Ollivier and W. H. Zurek,
		\href{https://link.aps.org/doi/10.1103/PhysRevLett.88.017901}
		{Phys. Rev. Lett. {\bf  88}, 017901 (2001)}.
		
		\bibitem{6.3LHenderson2001}
		L. Henderson and V. Vedral,
		\href{https://doi.org/10.1088/0305-4470/34/35/315}
		{J. Phys. A {\bf  34}, 6899 (2001)}.
		
		\bibitem{6.4PhysRevA.69.022309}
		M. Koashi and A. Winter,
		\href{https://link.aps.org/doi/10.1103/PhysRevA.69.022309}
		{Phys. Rev. A {\bf  69}, 022309 (2004)}.
		
		\bibitem{2024ATLAS}
		G. Aad et al. (ATLAS Collaboration),
		\href{https://doi.org/10.1038/s41586-024-07824-z}
		{Nature (London) {\bf 633}, 542 (2024)}.
		\bibitem{2024CMS1}
		A. Hayrapetyan et al. (CMS Collaboration),
		\href{https://doi.org/10.1088/1361-6633/ad7e4d}
		{Rep. Prog. Phys. {\bf 87}, 117801 (2024)}.
		
		\bibitem{2024CMS2}
		A. Hayrapetyan et al. (CMS Collaboration)
		\href{https://link.aps.org/doi/10.1103/PhysRevD.110.112016}
		{Phys. Rev.D {\bf 110}, 112016 (2024)}.
		\bibitem{CEPC2025}
		The CEPC Study Group,
		\href{https://arxiv.org/abs/2510.05260}
		{ arXiv 2205.08553v1{\bf}}.
		\bibitem{FCC-ee2019}
		A. Abada et al. (FCC Collaboration),
		\href{https://doi.org/10.1140/epjst/e2019-900045-4}
		{Eur. Phys. J. Special Topics {\bf  228}, 261 (2019)}.
		\bibitem{Kimble2008}
		H. J. Kimble,
		\href{https://doi.org/10.1038/nature07127}
		{Nature  (London) {\bf  453}, 1023 (2008)}.
		\bibitem{Zhong2021}
		Y. P. Zhong, H. S. Chang, A. Bienfait, É. Dumur, M. H. Chou, C. R. Conner, J. Grebel, R. G. Povey, H. X. Yan, D. I. Schuster, and A. N. Cleland
		\href{https://doi.org/10.1038/s41586-021-03288-7}
		{Nature  (London) {\bf  590}, 571 (2021)}.
		
	\end{references}
\end{document}